\documentclass[journal ]{new-aiaa}

\usepackage{graphicx}
\usepackage{amsmath}
\usepackage[version=4]{mhchem}
\usepackage{siunitx}
\usepackage{longtable,tabularx}
\usepackage[margin=1in]{geometry}
\usepackage{subcaption}
\usepackage{float}
\usepackage{hhline,multirow,float}
\makeatletter
\newcommand*{\Xbar}{}%
\DeclareRobustCommand*{\Xbar}{%
  \mathpalette\@Xbar{}%
}
\usepackage{amsmath}
\usepackage{amsfonts}
\usepackage{amssymb}
\usepackage{mathtools}
\DeclarePairedDelimiter{\ceil}{\lceil}{\rceil}
\usepackage{algorithm,algpseudocode}
\algnewcommand{\Inputs}[1]{%
  \State \textbf{Inputs:}
  \Statex \hspace*{\algorithmicindent}\parbox[t]{.8\linewidth}{\raggedright #1}
}
\algnewcommand{\Initialize}[1]{%
  \State \textbf{Initialize:}
  \Statex \hspace*{\algorithmicindent}\parbox[t]{.8\linewidth}{\raggedright #1}
}

\usepackage{tabulary,booktabs, threeparttable, stackengine,array,bm}
\usepackage{natbib,textcomp}
\usepackage{graphicx,xparse}
\usepackage[toc,page,title,titletoc]{appendix}
\usepackage{tikz}

\usepackage{bbold,algorithm,standalone}
\usepackage{algpseudocode}
\usepackage{wrapfig,enumitem, multirow}
\usepackage{tabularx} 
\usepackage{tabulary,booktabs, threeparttable, stackengine,array}
\usepackage{natbib,textcomp}
\usepackage{graphicx}
\usepackage[toc,page,title,titletoc]{appendix}
\usepackage{tikz}
\usetikzlibrary{shapes.geometric, arrows}
\definecolor{mygreen}{RGB}{0,100,0}
\tikzset{snake it/.style={decorate, decoration={snake, amplitude=0.3mm, segment length=2mm}}}
\tikzstyle{process} = [rectangle, minimum width=3cm, minimum height=1cm, text centered, draw=black]
\tikzstyle{decision} = [diamond, minimum width=3cm, minimum height=1cm, text centered, draw=black,aspect=4]
\tikzstyle{startstop} = [rectangle, rounded corners, minimum width=2cm, minimum height=1cm,text centered, draw=black]
\tikzstyle{io} = [trapezium, trapezium left angle=70, trapezium right angle=110, minimum width=3cm, minimum height=1cm, text centered, draw=black]
\tikzstyle{arrow} = [thick,->,>=stealth]
\tikzstyle{line} = [draw, thick, -latex']

\newcommand*\xbar[1]{%
  \hbox{%
    \vbox{%
      \hrule height 0.5pt 
      \kern0.5ex
      \hbox{%
        \kern-0.2em
        \ensuremath{#1}%
        \kern-0.1em
      }%
    }%
  }%
}
\newcommand{\figref}[1]{Figure \ref{#1}}
\newcolumntype{C}{>{\centering\arraybackslash}m{3cm}}
\newcolumntype{L}{>{\arraybackslash}m{2.5cm}}
\newtheorem{prop}{Proposition}
\newenvironment{proof1}
{\noindent {\bf Proof:}}{\vspace{1pc}}
 \bibpunct[, ]{(}{)}{,}{a}{}{,}%
\usepackage{float}
\usepackage{siunitx}

\title{Optimization of Distribution Network Configuration for Pediatric Vaccines using Chance Constraint Programming}

\author{Zahra Azadi}
\affil{University of Miami Herbert Business School, Department of Management.}
\author{Sandra D. Eksioglu}
\affil{University of Arkansas, Department of Industrial Engineering.}
\author{H. Neil Geismar}
\affil{Mays Business School, Department of Information and Operations Management.}

\begin{document}

\maketitle

\begin{abstract}
Millions of young people are not immunized in low- and middle-income (LMI) countries because of 
low vaccine availability resulting from inefficiencies in cold supply chains. 
We create supply chain network design and distribution models to address the unique characteristics and challenges facing vaccine supply chains in LMI countries. 
The models capture the uncertainties of demand for vaccinations and the resulting impacts on immunization, the unique challenges of vaccine administration (such as open vial wastage), the interactions between technological improvements of vaccines and immunizations, and the trade-offs between immunization coverage rates and available resources. The objective is to maximize both the percentage of fully immunized children and the vaccine availability in clinics.  Our research examines how these two metrics are affected by three factors: 
number of tiers in the supply chain, vaccine vial size, and  new vaccine technologies.
We tested the model using Niger's Expanded Program on Immunization, which is sponsored by the World Health Organization. We make many observations and  recommendations to help LMI countries increase their immunization coverage.
\end{abstract}

\section{Introduction}

This study develops stochastic models to optimize the cold supply chains for distribution of pediatric vaccines in low- and middle-income (LMI) countries.  We use these models to estimate how supply chain structure, vaccine vial size, and new vaccine technologies impact vaccine availability and, consequently, vaccine coverage rates in LMI countries.  This research represents a large step forward because, to our knowledge, all previous studies of this topic use either deterministic models or simulation.  We demonstrate how our stochastic model provides more accurate solutions that lead to higher percentages of fully immunized children (FIC), which is a key metric in this field.


Vaccines have been used for more than 50 years to prevent childhood diseases. In 1974, the World Health Organization (WHO) launched the Expanded Program on Immunization (EPI) with the goal of immunizing every child and pregnant woman around the world \citep{WHOBenefit}. The Global Vaccine Action Plan (GVAP), launched by the World Health Assembly in 2011, established a target for immunization coverage of 90\% for EPI vaccines by 2020 \citep{WHOImmunization}. Despite the success of GVAP over the last few years, in 2018 more than one in ten children worldwide did not receive vaccines for measles, diphtheria, or tetanus \citep{WHO20mill}. In Niger, the focus of our case study, the immunization coverage for the diphtheria-tetanus-pertussis (DTP) vaccine was 85\% in 2017, and 75\% in 2018 \citep{WHO20mill}.

We acknowledge that social, political, and economic factors lead to low vaccination coverage, but we focus on those factors within the supply chain, following \cite{KraiselYadav12}'s call to action for Supply Chain Management researchers to help improve global health supply chains for medicines, vaccines, and health technologies. Problems we consider include exposure to temperatures outside the required range of 2{\si\degree}C - 8{\si\degree}C, expiration, and physical damage.  Other logistical challenges that cause vaccine wastage in LMI countries are the lack of an adequate road network, the time required to reach rural health facilities, the vaccine overstock at the national central store arising from the inability of placing smaller orders more frequently, etc.  The case study of Section~\ref{sec-Results} reports a number of observations and provides recommendations for the development of policies and guidelines to facilitate the achievement of EPI targeted immunization coverage rates in Niger.

\subsection{Background on Vaccine Supply Chains in LMI Countries}
A typical EPI vaccine distribution network consists of four tiers: a national central store, regional stores, district stores, and clinics. United Nations Children's Fund (UNICEF) replenishes the inventory of the central store annually. Vaccines are shipped from the central store to the regional stores and from the regional stores to district stores using trucks. Refrigerated trucks are used for long distances.
The district stores use cars and motorcycles to ship vaccines to the clinics where children are vaccinated. The number of tiers in the supply chain might be greater or smaller than four, depending on the country. For example, Vietnam uses one additional tier between the regional and district stores. However, every supply chain has a central, national-level store where vaccines are inventoried for delivery to downstream supply chain members.


Cold supply chains with four or more tiers have lower transportation costs, but higher inventory processing/holding costs, than do supply chains with only three tiers. The smaller number of facilities in a three-tier supply chain leads to 
fewer broken vials (since vaccines are handled fewer times), compared to supply chains with four or more tiers. These trade-offs influence public health administrators' evaluation of how changes to the supply chain structure would impact costs and vaccine availability. 

\subsection{Research Questions}
The cold supply chain  plays a crucial role in delivering vaccines in LMI countries. These supply chains were established decades ago and were designed to deliver only six vaccines. During the last two decades, additional vaccines, including hepatitis B (HepB), hemophilus influenza type B (Hib), pneumococcal conjugate, rotavirus, and human papillomavirus were added to the schedule  \citep{WHO2018Facts}. With the expansion of immunization programs and the introduction of new vaccines, these supply chain structures have become outdated and strained beyond their capacity. For example,  a four-tier supply chain 
made sense 40 to 50 years ago because communication systems were underdeveloped. However, now a four-tier supply chain leads to vaccine overstock (because inventories are maintained in every store in every tier), which increases waste from damages, expiration, etc.
New and advanced communication channels enable public health administrators to track inventories in real-time, thereby rendering so many administrative layers unnecessary. Local health centers can now directly call the central stores to request inventories, bypassing regional or district stores \citep{Humphreys11}. 
This observation motivates our first research question: (R1) \emph{How does changing the number of tiers in the supply chain affect vaccine availability}?

Vaccines can be manufactured in a variety of vial sizes, typically 
1, 5, 10, or 20-doses (see Table~\ref{tab-VaccineChar}).  
We use the term \textit{vaccine packaging} to refer to the number of doses per vaccine vial.
A large vaccine vial with many doses uses less storage space per dose than does a smaller vaccine vial; thus, the corresponding inventory holding cost per dose is less. However 
a large-dose vaccine vial has more open vial wastage (OVW) than does a small-dose vaccine vial. 
These observations motivate our second research question: (R2) \emph{How does vaccine vial size affect the immunization coverage rates}?

\begin{table*}[h]
\centering \small \renewcommand{\arraystretch}{1.3}
\caption{Vaccine characteristics.} \label{tab-VaccineChar}
\resizebox{\textwidth}{!}{\begin{tabular}{|l|c|c|c|c|c|}
\hline
\bf{Vaccine}  & \bf{Nr. of doses} & \bf{Volume}  & \bf{Diluent volume} & \bf{Nr. of doses}  & \bf{Storage}\\
\bf{type} &  \bf{per vial} &  \bf{(\textit{cc/dose})} &  \bf{(\textit{cc/dose})} &  \bf{in regimen} &\\

\hline
BCG & 20 & 1.2 & 0.7 & 1& Refrigerator/freezer \\
\hline
Tetanus & 10 &3.0 & & 3&Refrigerator \\
\hline
Measles & 10 & 2.1 & 0.5 &2& Refrigerator\\
\hline
Oral Polio & 20 & 1.0 & & 4 & Freezer \\
\hline
Yellow Fever & 10 & 2.5 & 6.0 & 1 & Refrigerator/freezer \\
\hline
DTP-HepB-Hib & 1 & 16.8 & & 3 & Refrigerator\\
\hline
\end{tabular}}
\end{table*}


According to UNICEF, a pressing problem in LMI countries is vaccine shortages, which diminish the ability to meet vaccination demand. Many shortages occur because of the increased number and types of vaccines flowing in the supply chain: some countries have doubled the types of vaccines offered since the 1970s \citep{Humphreys11}. Larger populations also contribute to the increased volume in the supply chain. Furthermore, new vaccines have become bulkier. For example, dual-chamber vaccine injection devices are designed to simplify the reconstitution of vaccines before administration by pairing the vaccine and the diluent in the same unit. This alignment reduces wastage from OVW and other adverse events \citep{wedlock2018dual}, but dual-chamber devices are more expensive and use more storage space per dose than do the traditional EPI vaccines, which come in 10- and 20-dose vials \citep{wedlock2018dual}.

Most vaccines are stored at temperatures between 2\si{\degree}C and 8\si{\degree}C. The introduction of new vaccines plus the increased storage requirements per dose for dual-chamber devices require additional cold storage space. Countries that do not have the means to expand cold storage capacity postpone the introduction of new vaccines. A new technological advancement, thermostable vaccines, can resist heat, so they do not need refrigeration.   (The term  \textit{vaccine presentation} represents both dual-chamber and  thermostable vaccines.)   Such a technology increases the shelf life of vaccines without consuming cold storage space.  Hence, this innovation improves vaccine availability and reduces wastage. However, this new technology is expensive. These observations motivate our third research question: (R3) \emph{How does using new vaccine presentations affect immunization coverage rates}?

We develop a stochastic optimization model to address the research questions. The objective is to maximize the percentage of FIC and the vaccine availability in clinics, which is measured by the supply ratio (SR).  A fully immunized child receives every required dose of EPI vaccines specified by the vaccine regimen before turning five years old. The percentage of FIC measures the efficiency of the supply chain and of the public educational campaigns. 
SR measures only the efficiency of the supply chain because it represents  the percentage  of  children that  \emph{could}  be  vaccinated   using  the inventory available in all clinics. Thus, SR
is equivalent to \emph{inventory fill rate}  used in the supply chain literature. We use the term SR to facilitate comparison of our results to others' who study vaccine coverage.
Our model determines the need for storage and transportation capacity at each location, given the uncertain nature  of  the  demand  for  vaccinations.  We develop  a  case  study  using  real-life  data  from  Niger to evaluate the performance of the model. We make a number of observations, and we provide recommendations about vaccine distribution strategies and administration policies which lead to increases in FIC and SR.

\subsection{Research Contributions}
Only a few papers in the literature analyze the impacts of supply chain decisions on the immunization coverage rate of pediatric vaccines in LMI countries \citep{assi2013removing, chen2014planning, lee2017economic, wedlock2018dual, lee2012impact, azadi2019developing}. The existing research is limited in scope because either it conducts economic analyses of vaccination programs under different supply chain designs, or it proposes deterministic optimization models to maximize immunization coverage. Furthermore, previous models do not consider the compound effects of the numbers of tiers in the supply chain, the vaccine vial size, and the introduction of new vaccine technologies on vaccine availability at clinics. Our research  contributes to the literature in the following ways:
\begin{itemize}
\item We propose a \textit{data-driven, stochastic optimization} model that considers the random nature of demand for vaccines, which among other consequences  leads to high OVW, which reduces immunization coverage.
\item Our model facilitates the evaluation of how the \emph{number of tiers of the cold supply chain}, the \emph{vaccine vial size}, and the \emph{new vaccine technologies} impact immunization coverage rates.
\item 
Our model is validated using \textit{real-life data} from Niger. We present a number of observations and  make recommendations that will improve the performance of the country's cold supply chain.
\end{itemize}

This study is organized as follows: Section~\ref{sec-LitReview} presents a review of the relevant literature and identifies a research gap. We present the model and corresponding solution approach  in Section~\ref{sec-Model} and 
outline the case study from Niger  in Section~\ref{sec-CaseStudy}. Then, Section~\ref{sec-Results} summarizes the results and provides managerial insights. A summary of the results and suggestions for future work are in  Section~\ref{sec-Conclusion}.

\section{Literature Review} \label{sec-LitReview}
We first briefly summarize previous works on perishable inventory in general.  Then we narrow our focus to vaccine supply chains.

Most works on perishable inventory address food, especially fresh produce, as shown in many surveys \citep{Bakker12,Dave91,Goyal01,karaesmen2011managing,LiLanMaw10,NagurneyYu10,Nahmias82,Raafat}.  One outlier is \cite{Zhou11}, who optimize a periodic review inventory system for platelets used by hospital blood transfusion centers.  This is somewhat similar to our environment, being based in public health, but the study only considers one party, not the entire supply chain. 

We briefly discuss other studies that consider the larger perishable food supply chain.  \cite{BlackburnScudder} recommend a supply chain that is responsive from post-harvest to initial storage, then efficient beyond that.  \cite{Ketz15}  evaluate the use of sensors affixed to inventory to collect time and temperature information of perishables as they flow through a supply chain. \cite{Gaulkler20} address a similar environment to determine the optimal transportation mode for each link.  \cite{Govindan14} study distribution in a two-echelon location-routing problem with time-windows.  Their multi-objective model reduces traditional business costs plus those from greenhouse gas emissions.  A key distinction is that most studies of supply chains for perishable products have traditional business objectives of minimizing cost or maximizing revenues, which generally are secondary in public health applications.

\cite{DuijzerSurvey} survey the literature on vaccine supply chains.  Most such studies focus on supply chain problems for one of two groups of vaccines: ($i$)  seasonal vaccines, such as influenza, and ($ii$) nonseasonal ones, such as pediatric vaccines. Only a few  works in the literature propose models to design a cold supply chain network for influenza vaccines \citep{hovav2015network}. Some studies address the influenza vaccine's supply chain contracting problem \citep{chick2008supply,dai2016contracting}. {Other studies consider vaccine allocation decisions during pandemics \citep{duijzer2018dose,f1fbee18d195400980a3bc34a3f10f71}}. The focus of our research is nonseasonal vaccines.

A number of studies of nonseasonal vaccines consider the distribution network design for pediatric vaccines in LMI countries. Most of these works conduct an economic analysis of a proposed distribution network design and compare the results with conventional designs (see \cite{riewpaiboon2015optimizing}). Other works propose discrete event simulation models to evaluate the performance of specific vaccine distribution networks. Work by \cite{assi2011impact,assi2013removing}, \cite{haidari2013augmenting}, \cite{brown2014benefits}, \cite{haidari2015one}, \cite{lee2015landscaping, lee2016re, lee2017economic}, \cite{mueller2016impact},  and \cite{wedlock2018dual} use the Highly Extensible Resource for Modelling Event-Driven Supply Chains  (HERMES) software to study the supply chains in Niger, India, and Benin. For example, \cite{haidari2013augmenting} use HERMES to evaluate how adding stationary storage  and transportation capacity impacts vaccine availability at clinics in Niger. Several other papers use  HERMES to explore the impact of different vaccine distribution strategies on supply chain costs and vaccine availability in clinics \citep{assi2013removing,brown2014benefits,lee2015landscaping,haidari2015one}.    
In contrast, we propose a stochastic optimization model 
for supply chain decisions under uncertainty. 

Some papers use HERMES to explore the impact of vaccine packaging on supply chain costs and vaccine availability at clinics), and other papers use Monte Carlo simulation \citep{dhamodharan2012determining}. \cite{assi2011impact} simulate the supply chain of EPI vaccines in Niger. The authors analyze the performance of Niger's vaccine distribution network if the measles vaccine vial size were to change from 10-dose to 5-, 2-, or 1-dose. A recent study by \cite{azadi2019developing} evaluates the trade-offs between using a single, fixed vial size for each vaccine and using a combination of vial sizes. \cite{azadi2019developing}  assume a two-tier supply chain consisting of the central store and the clinics.  Clinics observe the daily stochastic demand and decide when and which vaccine vial to open to minimize OVW. By focusing on optimizing the daily operational decisions of inventory management and vaccine administration, they conclude that using a combination of 1- and multi-dose vials leads to improved performance of the cold supply chain and, thus, to increased vaccine availability.
Our models  extend the stochastic optimization model developed  by \cite{azadi2019developing} by evaluating the impact of strategic decisions (i.e., changing the number of tiers in the supply chain) and of planning decisions (i.e., changing vaccine packaging) on the level of FIC and of SR in Niger.

\cite{lee2017economic} use HERMES to evaluate the economic impacts of thermostable vaccines, and \cite{wedlock2018dual} use HERMES to examine how the measles-rubella vaccine's dual-chamber injection device effects OVW and vaccine availability. The structure of the supply chain is fixed in both studies (as it is for all that use HERMES), and vaccines are pushed down the supply chain to meet demand. Our optimization model, in contrast, can specify an optimal redistribution among all vaccines of the resources that become available when a new vaccine presentation is implemented for a subset of vaccines. Furthermore, in addition to maximizing vaccine availability SR, our objective maximizes the percentage of FIC. Maximizing the percentage of FIC requires clinics to maintain a balanced inventory of vaccines, which influences decisions in the supply chain. 

\cite{chen2014planning} use the estimated mean demand for vaccines to evaluate the performance of three- and four-tier supply chains. 
Relying solely on the mean demand can lead to inefficient decisions in the cold supply chain and may reduce the study's realism. For example, \cite{chen2014planning}  found that the performance of three- and four-tier supply chains of Niger does not impact the number of FIC. Our numerical analysis  indicates that the three-tier supply chain leads to higher FIC than does the four-tier supply chain. This finding through stochastic optimization provides further evidence of the observation \cite{assi2013removing} made through simulation.

\section{Model Description and Solution Approach} \label{sec-Model}
This section establishes a modeling framework for the cold supply chain that delivers vaccines from an LMI country's point of entry to the clinics that administer vaccines to children. 
The objective is to maximize the percentage of FIC by increasing vaccine availability  at clinics. The availability of vaccines is represented by SR, the proportion of demand that is met from the available inventory. The network design of the cold supply chain, the vaccine vial size, and the vaccine presentation impact vaccine availability, hence we combine FIC and SR to form the appropriate objective to answer our research questions. 

This section begins by presenting the assumptions, which are motivated  by practices in the field. After describing the conceptual framework, we present the mathematical model.

\subsection{Assumptions}
The WHO and other nonprofit organizations deliver pediatric vaccines to LMI countries. These organizations work closely with the countries' governmental officials to make manufacturer selections and procurement decisions.   We assume that these 
decisions have been made so that we can focus on a receiving country's cold supply chain network, which begins at the nation's central store (where vaccines are received) and ends at the clinics around the country.  

Practices in Niger and the standards of the EPI cold supply chain inspire the following  assumptions:
\begin{itemize}
    \item The cold supply chain delivers only EPI vaccines.
    \item The planning horizon is one year, and the planning period is one month.
    \item The vaccine vial size represents the number of doses per vial. 
    Vial size is fixed in our analysis of research questions R1 and R3. 
    \item Manufacturing capacity is not a bottleneck.
    \item Since vaccines have different characteristics,  some can be stored in refrigerators and freezers, such as Bacillus Calmette–Guérin (BCG), and others must be stored in refrigerators, like the measles vaccine.
    \item Vaccines that are freeze-dried before shipping are restored prior to injection with a diluent. The diluent is refrigerated the day before reconstitution so that it is the same temperature as the freeze-dried vaccine.
    \item Some of the EPI vaccines require one dose for complete immunization, and others require multiple doses at different stages of childhood, such as the measles vaccine. 
    The number of doses required to immunize children fully is the \textit{vaccine regimen}.
    \item The expected OVW in clinics depends on the expected number of patient arrivals and on the vaccine vial sizes. OVW rates are calculated offline using this information.
    \item  The demand for vaccines in storage facilities is zero.
    \item The existing cold supply chain has limited refrigeration and freezer capacity.
    \item  The supply chain has the required transportation capacity to deliver vaccines.
\end{itemize}

\subsection{Conceptual Framework}
\label{ssec:CF}
This section presents the structure of the mathematical model and describes the constraints intuitively. 

The goal of the optimization model is to maximize the percentage of FIC and the vaccine availability, measured via SR, in LMI countries during the planning horizon. 
Considering both objectives provides a clearer picture of the EPI's effectiveness. 

The decision variables are the  number of vaccine vials (and their sizes, if appropriate) shipped from an upstream store to a downstream store or clinic. 
These determine the inventory of vaccines stored at each location, the amount of demand satisfied at each clinic,  and the number of FIC at each clinic.

The constraints are categorized as follows.
\begin{enumerate}
    \item \textbf{Inventory balance constraints}: The number of vaccine vials in a store in period $t$ depends on the store's inventory from the previous period and on the amount shipped (both in and out) during period $t$.
    \item \textbf{Capacity constraints}: The number of vaccines inventoried in a store is limited by the number and the capacity of its refrigerators and freezers.
    \item \textbf{Inventory initialization constraints}: The inventory of  vaccine vials in stores at the beginning of the planning horizon is zero.
    \item \textbf{Storage differentiation constraints}: 
    The storage space for a vaccine depends on the vial size and the vaccine type.
    \item \textbf{FIC constraints}: The number of FIC is limited by the number of vaccines available at the clinics.
    \item \textbf{Demand constraints}: The probability of vaccinating every child who visits a clinic  must be higher than a specified threshold level.
\end{enumerate}

Whereas the decision to use a three- or four-tier supply chain is a  strategic decision, inventory replenishment and transportation scheduling are operational decisions. Our model integrates these different types of decisions under uncertainty. The demand for vaccination at a clinic is modeled as a stochastic parameter, and
the demand constraint is modeled as a chance constraint. 
We use the comprehensive data analysis in \cite{azadi2018iise} to develop the distribution of demand.

We model both a four-tier supply chain and a three-tier supply chain for distribution of vaccines.  
Both models are then updated to consider the implications of using different vaccine vial sizes and of using a new technology that impacts vaccine presentation.
Both models are stochastic linear programs that maximize the percentage of FIC and the vaccine availability SR.

\subsection{Mathematical Model}
Our model increases vaccine availability by identifying the appropriate number of tiers in the supply chain and by developing a plan for both inventory replenishment and transportation of vaccines.


Let $G_t=(\mathcal{J}, \mathcal{A})$ be a directed graph that represents the vaccine supply chain network at time period $t$ during a planning horizon of length $\mathcal{T}$ $(t \in \mathcal{T})$.  In this graph, $\mathcal{J}$ is the set of nodes, and~$\mathcal{A}$ is the set of arcs. Let $J (\subset \mathcal{J})$ represent the set of clinics and $\mathcal{I}$ represent the set of vaccine types. Let ${\Delta}_{ijt}$ represent the stochastic demand for vaccine $i \in \mathcal{I}$
at a clinic $j \in {J}$ during  period $t \in \mathcal{T}$, and let $\mu_{ijt}$ represent the corresponding expected value. Note that ${\Delta}_{ijt}$ follows a Poisson distribution because it represents the number of patients arriving at a clinic in period $t$. The distribution of~${\Delta}_{ijt}$ can be approximated with a Normal distribution $\bar{\Delta}_{ijt} $ because
$\mu_{ijt}> 10$ 
\citep{devore2011probability}. Since demand for vaccinations is positive, we approximate it using a truncated  Normal random variable ${\Delta}_{ijt}\approx \max\{0, \bar{\Delta}_{ijt}\}$.

Let $x^{R}_{ijt}$ and $x^{F}_{ijt}$ represent the number of vaccines refrigerated and frozen at a clinic in period $t$. Meeting all demand for vaccinations would require 
\begin{equation}
 x^{R}_{ijt}+x^{F}_{ijt} \geq \Delta_{ijt},  \quad \quad \forall i \in \mathcal{I}, j \in J,t \in \mathcal{T}. \label{Con-DemandLeq}
\end{equation}
This deterministic constraint cannot always be met in LMI countries, given their capacity limitations and the cost of maintaining a high inventory level.
Thus, these constraints are softened by requiring that they be satisfied for only a specified percentage of combinations of $i$, $j$, and~$t$. Let $\beta$ be the risk parameter selected by clinic $j \in {J}$ for vaccine $i \in \mathcal{I}$.
The nominal value of this parameter, $10\%$,  is motivated by Global Alliance for Vaccines and Immunizations'  target immunization coverage of 90\% for EPI vaccines \citep{WHOImmunization}.  We update constraints~\eqref{Con-DemandLeq}  using  chance constraints~\eqref{Con-DemandLeqChanceCons}, which  indicate that the number of vaccines available in every time period should be greater than or equal to demand in at least $(1-\beta)\%$ of cases:
\begin{equation}
\bm{P} \bigg(x^R_{ijt}+x^F_{ijt} \geq {\Delta}_{ijt} \bigg)\geq (1-\beta) \quad \quad \forall i \in \mathcal{I}, j \in J,t \in \mathcal{T}. \label{Con-DemandLeqChanceCons}
\end{equation}

Chance Constraint Programming (CCP)  formulates an optimization problem with a probabilistic constraint.  CCP has been used to model optimization problems which arise in a number of fields, such as healthcare \citep{beraldi2004designing}, financial risk management \citep{ donkor2013optimal}, and renewable energy generation \citep{lubin2016robust},  among others. A  complete CCP model formulation, which we refer to as formulation $(P)$, is presented in Appendix \ref{appA}, along with a list of the decision variables and the parameters.

The following succinct formulation of  $(P)$ simplifies the exposition of the algorithm presented in \S \ref{sec-SolMeth}. This formulation uses a single decision vector $\bf{x} \in \mathcal{X}$ to denote collectively all the decision variables described in Appendix \ref{appA}. The objective function, $F({\bf x})$, maximizes the percentage of FIC and the vaccine availability SR by summing these two with a weighting factor. To find the number of FIC, we calculate the number of children who received every dose specified by the vaccine regimen. Vaccine availability represents the total number of doses available in clinics. Constraints~\eqref{Con-RefrigInv}-\eqref{Con_nonnegativity} in Appendix \ref{appA}, summarized in \S\ref{ssec:CF}, define the feasible region $\mathcal{X}$:
\begin{subequations} \label{model-general}
\begin{align}
(P): \quad\quad  & \max_{\bf{x} \in \mathcal{X}} ~ F({\bf x})  \\ \quad\quad s.t.:\quad\quad
    &\bm{P}\bigg(G_{ijt}(\textbf{x},{\Delta}_{ijt}) \geq 0\bigg) \geq (1-\beta ), \quad \quad \forall i \in \mathcal{I}, j \in {J}, t \in \mathcal{T},\label{model-generalCS} \\
\text{where} \notag\\
& G_{ijt}(\textbf{x},{\Delta}_{ijt})= x^R_{ijt}+x^F_{ijt} - {\Delta}_{ijt}. \notag
\end{align}
\end{subequations}

$F({\bf{x}})$ is a continuous real valued function, and function $G_{ijt}(\textbf{x},{\Delta}_{ijt})$, which represents either unmet demand or excess inventory in period $t$, is measurable for every $\bf{x} \in \mathbb{R}^n$ and continuous for  every feasible ${\Delta}_{ijt}$. 
Let $\upsilon^*$ be the optimal objective value of (P). We assume that $\upsilon^*$ is bounded and that the feasible region $\mathcal{X}$ is non-empty.

Our model does not have budget constraints, but it does consider transportation and storage capacity  constraints. This is in line with other models of humanitarian supply chains that focus on improving human lives and well-being, rather than on minimizing costs or maximizing profits \citep{beamon2008performance, mccoy2014using, balcik2015measuring}. 

\subsection{Solution Approach} \label{sec-SolMeth}
CCP  models are difficult to solve because of ($i$) the non-convexity of the feasible set defined by chance constraints, and ($ii$) the inability to compute the probability of satisfying chance constraints (arising from the computational challenge of multi-dimensional integration) \citep{van2011chance, kim2015guide}. However, there are a few exceptions to this observation. For example, researchers have developed computationally tractable approximations of CCP models with single chance constraint(s) when the distribution of uncertain parameters belongs to a family of \textit{logarithmically concave distributions} \citep{kuccukyavuz2012mixing, geletu2013advances}.%
\footnote{The interest reader may consult \cite{nemirovski2012safe}'s survey of scenario-based and computationally tractable approximations of chance constraints.}  In general, the development of a tractable  approximation to a chance constraint is not an option \citep{henrion2004introduction}. This is why many researchers use the sample average approximation (SAA) method to solve CCP models.

SAA is a scenario-based approach to solving stochastic optimization problems. The scenarios considered are generated via Monte Carlo simulation.  
Each scenario represents a realization of problem parameters drawn from 
empirical distributions. 
These realizations are used to develop a corresponding deterministic equivalent formulation (DEF) \citep{pagnoncelli2009sample, jiang2016data}. The DEF guarantees that the number of instances that do not satisfy constraints~\eqref{Con-DemandLeq} in the independent trials of random sampling is below a certain level. 

Because the random parameter 
${\Delta}_{ijt}$ follows a truncated Normal distribution, a closed-form representation of the chance constraints cannot be derived. 
Instead, SAA  is used to solve (P).

{\bf An SAA of Chance Constraints}: In the SAA method, the true continuous probability function $\bm{P}$ in \eqref{model-generalCS} is approximated by its empirical counterpart $\bm{P}^S$. To develop $\bm{P}^S$, we begin by generating 
a sample  $\{\bm{\delta}_{ijt}^1, \bm{\delta}_{ijt}^2, \dots, \bm{\delta}_{ijt}^S\}$ of $S$  independent, identically distributed (iid)  realizations of the uncertain parameter ${\Delta}_{ijt}$.  Let $p_s = 1/S,~s=1, \dots,S$ be the probability of each realization. Let $\Lambda(\bm{\delta}_{ijt}^s)$ represent a measure of mass 1 at point $\bm{\delta}_{ijt}^s$, then, $\bm{P}^S:=\sum_{s=1}^{S}p_s\Lambda(\bm{\delta}_{ijt}^s)$ \citep{pagnoncelli2009sample}.

Let $ \mathbb{1}_{(0,\infty)}(.):\mathbb{R} \xrightarrow{} \mathbb{R}$ be an indicator function 
such that $ \mathbb{1}_{(0,\infty)}(\ell):=1$ if $\ell >0$, and $ \mathbb{1}_{(0,\infty)}(\ell):=0$ if $\ell \leq 0$. Using this function we derive Problem~($\hat{P}$), which is the deterministic SAA formulation of ($P$):
\begin{subequations}
\begin{align}
(\hat{P}): \quad\quad \max_{\bf{x} \in \mathcal{X}} ~ &F({\bf x}) \label{model-general-SAA-obj} \\
    s.t. \hspace{0.2in}
& \frac{1}{S}\sum_{s=1}^{S}(\mathbb{1}_{(0,\infty)}(G_{ijt}(\textbf{x},\delta^s_{ijt}))) \geq (1-\hat{\beta} ), \quad \quad \forall i \in \mathcal{I}, j \in J, t \in \mathcal{T},\label{model-general-SAA-con}
\end{align}
\end{subequations}
where $\hat{\beta} \geq 0$ is 
different from the risk factor ($\beta$) of the original problem. Let $\upsilon^S$ be the optimal solution to Problem~($\hat{P}$) and $\mathcal{X}^S$ be the corresponding feasible region. \cite{luedtke2008sample} show that if $1-\hat{\beta} >1-\beta$, then every feasible solution of Problem~($\hat{P}$) is with high probability also feasible for Problem (${P}$) as the number of realizations $S$  increases. Stated formally, $\upsilon^S \xrightarrow{} \upsilon^* ~~ \text{and} ~~ \mathcal{X}^S \xrightarrow{} \mathcal{X} ~~ w.p.1 ~~ \text{as} ~~ S \xrightarrow{} \infty.$

Constraints \eqref{model-general-SAA-con} use of an indicator function renders them computationally intractable by commercial solvers. We reformulate these constraints by introducing the continuous variables~${v}_{ijt}$ and~$w_{ijt}$, and the penalty term $\pi_{ijt}$. Variables ${v}_{ijt}$ and $z_{ijt}$ are slack variables that measure violations of constraints \eqref{model-general-SAA-con}. A ${v}_{ijt} > 0$ indicates a shortage, and a $z_{ijt} >0$ indicates excess of vaccine type $i$ at location~$j$ in period $t$. The parameter $\pi_{ijt}$ represents a penalty for not serving demand in time $t$.  The resulting reformulation of ($\hat{P}$) is
\begin{subequations}
\begin{align}
(\bar{P}):\quad\quad  \max_{\bf{x} \in \mathcal{X}} & \quad F({\bf x})-\sum_{s=1}^{S}\big(\sum_{i \in \mathcal{I}}\sum_{j \in J}\sum_{t \in \mathcal{T}}\pi_{ijt}{v}^s_{ijt}\big) \quad\quad \\s.t.\quad
&  G_{ijt}(\textbf{x},\delta^s_{ijt})+{v}^s_{ijt}-z^s_{ijt}=0,\quad \quad \forall i \in \mathcal{I}, j \in J, t \in \mathcal{T}, s=1,\dots,S, \label{CahcneContraint-Reformulated-SAA}\\
  &  {v}^s_{ijt}, z^s_{ijt} \geq 0, \quad \quad \quad \quad \quad \quad \quad \quad \quad  \forall i \in \mathcal{I}, j \in J, t \in \mathcal{T}, s=1,\dots,S. \label{SlackTypes-SAA}
\end{align}
\end{subequations}

The values of  $\pi_{ijt}$ are not known in advance.  Instead, they are adjusted by Algorithm Binary Search Sample Average Approximation (BS-SAA) in Appendix~\ref{appB} so that $v^s_{ijt} = 0$ in at least $\ceil {(1-\hat{\beta})S}$ of the scenarios, as required by constraints~\eqref{model-generalCS}.
If a specific $\pi_{ijt}$ is too high, then the maximization problem would force $v^s_{ijt} = 0$, for all $s=1,\dots,S$, and consequently
$G_{ijt}(\textbf{x},\delta^s_{ijt}) \geq 0$ for all $s=1,\dots,S$. 
\begin{prop}
Algorithm BS-SAA generates a feasible lower bound for problem ($P$).
\end{prop}
\begin{proof1}
Algorithm BS-SAA iteratively changes the values of $\pi_{ijt}$, solves ($\bar{P}$), and counts the number of scenarios $s\in \{1,2,\ldots,S\}$ for which $v^s_{ijt} = 0$. The algorithm ends when $v^s_{ijt} = 0$ in at least $\ceil {(1-\hat{\beta})S} + \epsilon$ of the scenarios generated, where $\epsilon > 0$ 
(see steps (11) and (13) of the BS-SAA algorithm).

 The solution obtained by the algorithm approaches the optimal solution of ($P$) as $\epsilon  \xrightarrow{} 0$ and $S \xrightarrow{} \infty$ \citep{luedtke2008sample}. 
 Thus, Algorithm BS-SAA  provides a lower bound for model~($P$). $\blacksquare$
\end{proof1}

\section{The Case Study and the Experimental Setup} \label{sec-CaseStudy}
We develop a case study using data from Niger. We first introduce the data sources and describe the structure of the distribution network. Following that, we detail our experimental setup, which includes a scenario selection procedure and an evaluation of the stochastic solution's  performance.

\subsection{The Case Study: Vaccine Supply Chain of Niger}
\begin{figure}
    \centering
    \includegraphics[scale = 0.4]{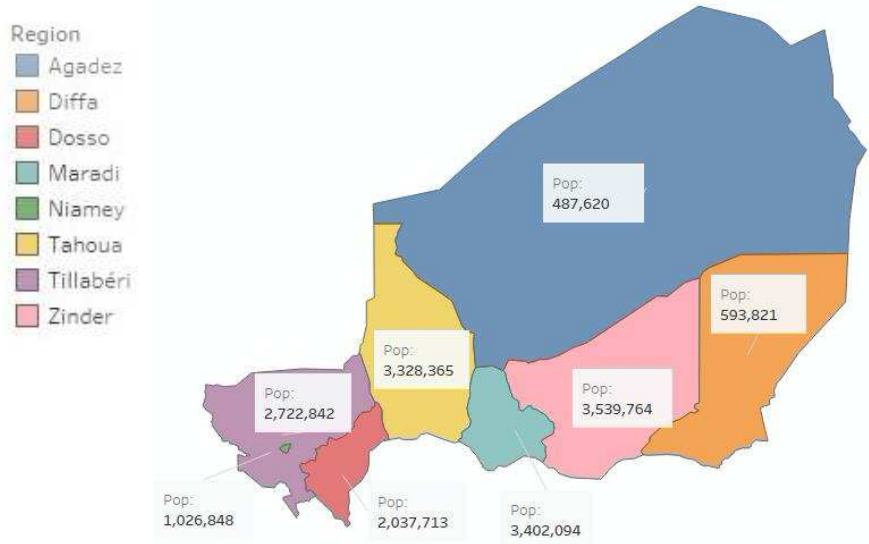}
  	\caption{Niger map with major demand zones.}
    \label{fig:nigermap}
\end{figure}

\figref{fig:nigermap} presents a map of Niger's major demand areas. Most of the population in Niger resides in the South, near the border with Nigeria. We identified eight demand regions: Niamey, Tillaberi, Dosso, Maradi, Tahoua, Agadez, Zinder, and Diffa. \cite{azadi2018iise}'s data analysis of the demand for vaccines in these regions makes two important observations: ($i$) population size, birth rates, and  adult literacy rates can predict the expected demand for vaccines, and ($ii$) the demand for vaccines differs among Niger's different regions. Thus, we use the region-based linear regression models developed by \cite{azadi2018iise} to determine the expected monthly demand for vaccinations in each clinic. These regression models are fitted to the available data on childhood vaccinations in the different regions of Niger from 2008 to 2012. We use population size as the independent variable to estimate the expected monthly demand for vaccination in each region. These data are obtained from the Demographic and Health Surveys (DHS) \citep{DHS}. Since the  data are aggregated at the regional level, we divide each region's demand equally among the clinics of the region.

The regression functions are of the form $\bar{\Delta}_{ijt} = \mu_{ijt} + {\omega}_{ijt}$, where ${\omega}_{ijt}$ represents the residual term. This residual follows a Normal distribution with mean 0 and constant standard deviation, $\sigma_{ijt}$. As a result, the $\bar{\Delta}_{ijt}$ 
follow a  Normal distribution, $N(\mu_{ijt},\sigma_{ijt})$. Our numerical experiments only consider $\bar{\Delta}_{ijt} > 0$; thus, we use ${\Delta}_{ijt}= \max\{0,\bar{\Delta}_{ijt}\}$, which follows the corresponding truncated Normal distribution.

\begin{figure}
    \centering
    \includegraphics[width=\textwidth]{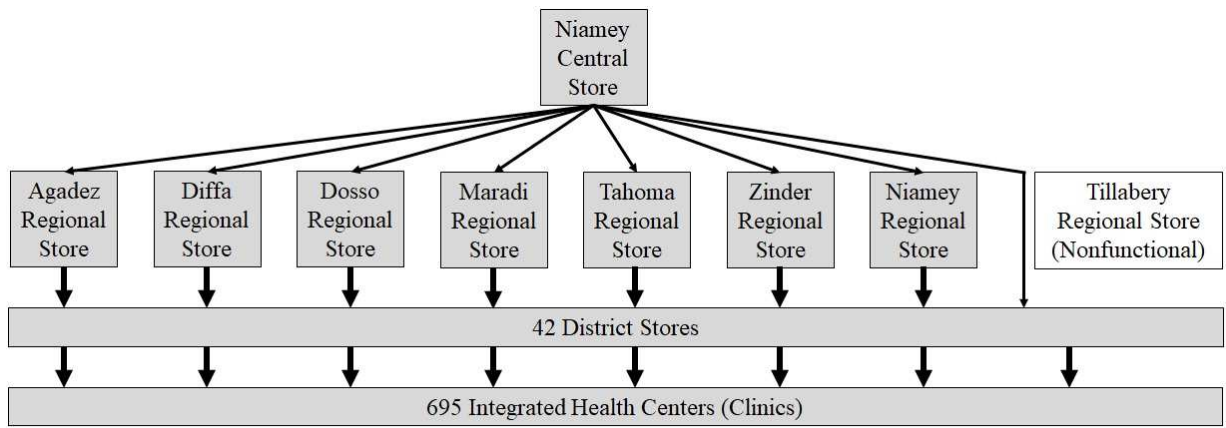}
  	\caption{Vaccine supply chain in Niger. Adapted from \cite{chen2014planning} }
    \label{fig:NigerVSC}
\end{figure}

The vaccine distribution network in Niger consists of four tiers (See \figref{fig:NigerVSC}). The government of Niger purchases vaccines from the UNICEF headquarters in Denmark, and UNICEF uses planes to replenish the inventory of the central store at Niamey every two months.  Vaccines are shipped from the central store to the regional stores in Niamey, Tillaberi, Dosso, Maradi, Tahoua, Agadez, Zinder, and Diffa every three months via cold trucks. The district stores pick up vaccines from the regional stores every month via $4 \times 4$ trucks. The regional store in Tillaberi is currently non-functional, so the district stores in this region receive vaccines directly from the central store at Niamey. The  clinics  pick up vaccines from the district stores monthly using motorcycles, bicycles, or private cars. This network consists of eight regional stores, 42 district stores, and 695 clinics. The central and regional stores are equipped with cold rooms. The district stores use chest refrigerators and freezers, and the clinics use smaller refrigerators and/or freezers. The set of EPI vaccines currently administered in Niger  are BCG, tetanus, measles, oral polio, yellow fever, and DTP. The characteristics of each vaccine are summarized in Table \ref{tab-VaccineChar}  \citep{chen2014planning}. We consider a planning horizon of 12 months.


\subsection{Experimental Setup} \label{sec-CompResults}
The BS-SAA algorithm is developed using JuMP programming language. The large-scale linear program for model ($\bar{P}$) is solved using Gurobi Solver version 7.0.2. The experiments are implemented using the high-performance computing  resource at Clemson University \citep{Palmetto}.

The literature  measures immunization coverage rates by SR$_{is}$, which represents the average SR for vaccine type $i$ under scenario $s$, and by FIC$_{js}$, which represents the average percentage of fully immunized children in region $j$ under scenario $s$ \citep{chen2014planning}. 
These measures are calculated as
\begin{subequations}\label{eq-performanceMeasures}
\begin{eqnarray}
& SR_{is}=\dfrac{\sum_{j \in J}\sum_{t \in \mathcal{T}}(x^R_{ijt}+x^F_{ijt})}{ \sum_{j \in J}\sum_{t \in \mathcal{T}} (\Delta_{ijt}^s)} \label{eq-metricSR},  \quad \quad \forall i \in \mathcal{I}, s=1,\dots,S, \label{eq-SupplyRatio}\\
\notag\\
& {FIC}_{js} = 100 \dfrac{n_{j}} { \sum_{i \in \mathcal{I}}\sum_{t \in \mathcal{T}} (\Delta_{ijt}^s)}    \label{eq-metricFIC}, \quad \quad \forall j \in J, s=1,\dots,S, \label{eq-FICRatio}
\end{eqnarray}
\end{subequations}
where 
$n_{j}$ is the total number of fully immunized children at clinic $j$,  $n_j \leq \sum_{t \in \mathcal{T}}(x^R_{ijt}+x^F_{ijt})/a_i, ~ \forall i\in \mathcal{I}$; and $a_i$ is the number of doses of vaccine $i$ specified by the vaccine regimen.

\subsection{Evaluating the BS-SAA Algorithm}
We use the approach proposed in \cite{luedtke2008sample} to identify the sample size $S$ and the number of replications $M$ to use in our experimental analysis. This approach ensures with high confidence that a solution to model ($\bar{P}$) is also a feasible solution of high quality to the original problem ($P$).  \cite{luedtke2008sample} show that the confidence levels for our results are at least $1 - \left(\frac{1}{2}\right)^M$, so we choose  $M=10$, which yields  $1 - \left(\frac{1}{2}\right)^{10} = 0.999$. 

\medskip

\noindent {\bf Finding a feasible solution (lower bound (LB)):}  
We solve ten iterations ($\bar{P}$) using a sample of size $S$. Let ${\bf \overline{x}}$ be a feasible solution to ($\bar{P}$), and let $\overline{SR}$, $\overline{FIC}$ be the corresponding values of SR and FIC. Next, we conduct a posterior check using a large sample of size $S^{\prime}$. Our posterior check uses this new sample of $\Delta_{ijt}^s$ $(s\leq S^{\prime})$ to recalculate SR and FIC. These values are compared  to  $\overline{SR}$ and $\overline{FIC}$. If the difference found is not significant, then we consider ${\bf \overline{x}}$ to be a feasible solution to  problem ($P$).

Because SR measures the percentage of children that can be vaccinated in a clinic using the available inventory, SR is the probability that demand for vaccination is met, which is equivalent to the chance constraints. Hence our posterior check focuses on SR. 

We solve model ($\bar{P}$) for a four-tier supply chain using samples of size $S =$ 10, 30, 50, $\ldots$  300. In these experiments,  $\beta = 0.80$ and $\hat{\beta} = 0.70$. Note that we could not find feasible solutions for smaller values of $\hat{\beta}$ because of the limited storage capacity.  We use S$^{\prime} = 300$ for the posterior analysis. Table \ref{tab-Sol} summarizes the results of our experiments.  All replications generated feasible solutions.  We present  the minimum, maximum, and average FIC and SR values of the solutions, and the average running time.  Because increasing the sample size from 250 to 300 has only a marginal impact on the quality of the solutions, we set  $S = 300$ in the remaining experiments.

\begin{table}[h]
\caption{Evaluating the Quality of Feasible Solutions ($\hat{\beta} = 0.70$)} \label{tab-Sol}
\centering
\begin{tabular}{rccccccc}
\hline
  &\multicolumn{3}{c}{\bf Average SR (\%)} & \multicolumn{3}{c}{\bf Average FIC (\%) } & {\bf Avg Run}\\
  \cline{2-4}\cline{5-8}
{\bf S}   & {\bf Min}  & {\bf Max} & {\bf Avg} & {\bf Min}  & {\bf Max} & {\bf Avg}& {\bf Time (sec)}\\ 
\hline
 10 &  36.68 & 37.33 & 36.95 & 11.67 & 12.17 & 12.03 & 97 \\
30 &   32.81 & 33.17 & 33.03 & 10.08 & 10.72 & 10.48  & 146\\
50 &  30.40 & 31.03 & 30.74 & 8.76 & 9.24 & 9.04 & 163\\
70 &  28.28 & 28.83 & 28.58 &  8.05 & 8.41 & 8.32 & 133\\
100 &   26.51 & 26.88 & 26.66 & 7.56 & 7.80  & 7.73 & 164 \\
200 &  24.03 & 24.17 & 24.10 & 6.86 & 6.99 & 6.92 & 202\\
250 &  23.44 & 23.64 & 23.55 & 6.71 & 6.81 & 6.75 & 289\\
300 &  23.10 & 23.24 & 23.17 & 6.59 & 6.69 & 6.63 & 776\\
\hline
  \hline
\end{tabular}
\end{table}%

\noindent {\bf Finding an upper bound (UB):} 
To evaluate the quality of our solutions, we solve model ($\bar{P}$) for a four-tier supply chain, using $S =300$ and $\hat{\beta}=\beta = 0.80$. We find the maximum SR and maximum FIC values across $M=10$ replications. We consider these values to be upper bounds for SR and FIC. Next, we calculate the percentage difference (error gap) between these values and the feasible solutions presented in Table \ref{tab-Sol} for $S$=300.  Table \ref{tab-Error} demonstrates that our algorithm results are quite good.

\begin{table}[h]
\caption{Evaluating the Upper Bounds ($\hat{\beta} = 0.80$)} \label{tab-Error}
\centering
\begin{tabular}{lcccccc}
\hline
  & \multicolumn{3}{c}{\bf Average SR (\%)} & \multicolumn{3}{c}{\bf  Average FIC (\%) } \\
  \cline{2-4}\cline{5-7}
{\bf S}   & {\bf Min}  & {\bf Max} & {\bf Avg} & {\bf Min}  & {\bf Max} & {\bf Avg}\\ 
\hline
300 & 2.98 & 3.57 & 3.28 & 8.76 & 10.19 & 9.58 \\
\hline
  \hline
\end{tabular}
\end{table}%

\section{Discussion of Results} \label{sec-Results}
\begin{figure}[h] 
\centering
  \begin{subfigure}{0.5\textwidth}
        \centering
 	    		\includegraphics[width=\textwidth]{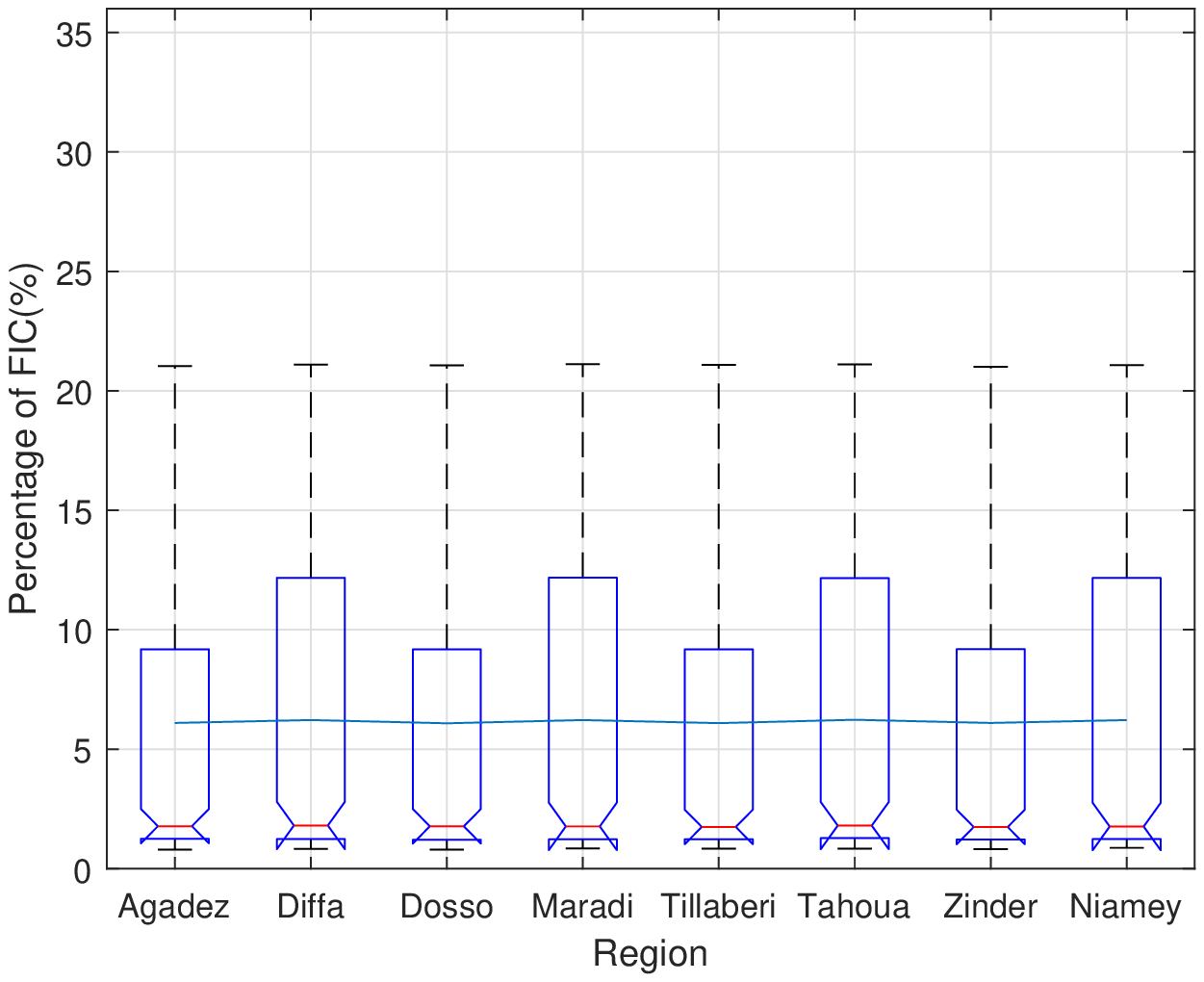}
  	\caption{Four-Tier Supply Chain}
	\label{fig:FIC_Base_4Tiers}
    \end{subfigure}
    \begin{subfigure}{0.5\textwidth}
        \centering
 	   	\includegraphics[width=\textwidth]{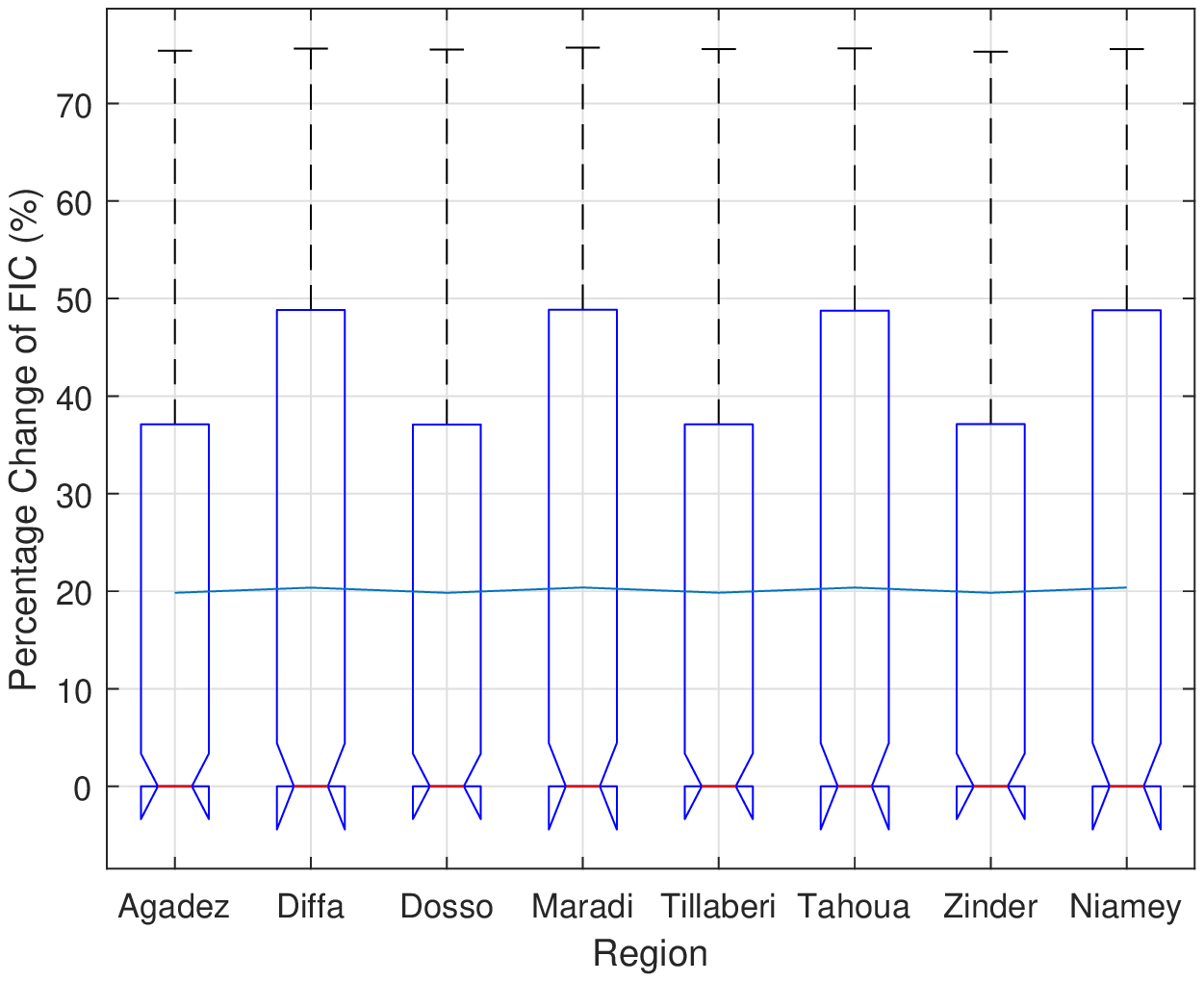}
   	\caption{Capacity Added to Clinics}
	\label{fig:FIC_Base_3TiersIHC}
    \end{subfigure}%
    \begin{subfigure}{0.5\textwidth}
        \centering
 	    		\includegraphics[width=\textwidth]{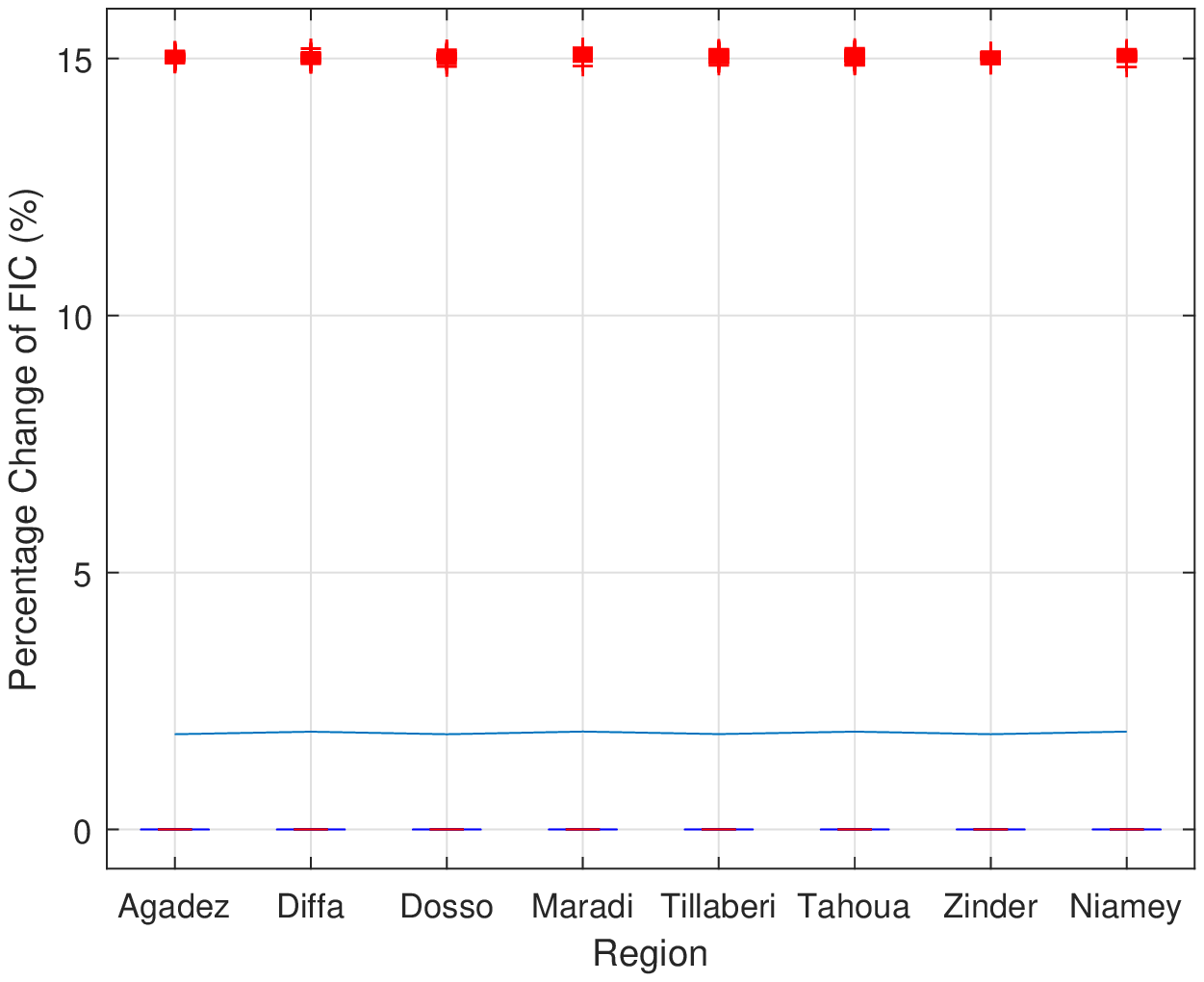}
  	\caption{Capacity Added to  District Stores}
	\label{fig:FIC_Base_3TiersDist}
    \end{subfigure}

    \caption{(a) Region-Based FIC for Four-Tier Supply Chain;  (b)  \% Increase in FIC in a Three-Tier Supply Chain if capacity is added to Clinics; (c) \% Increase in FIC in a Three-Tier Supply Chain if capacity is added to District Stores.} \label{fig::FICRegions_4and3}
\end{figure}

\begin{figure}[h] 
\centering
    \begin{subfigure}{0.5\textwidth}
        \centering 	    		\includegraphics[width=\textwidth]{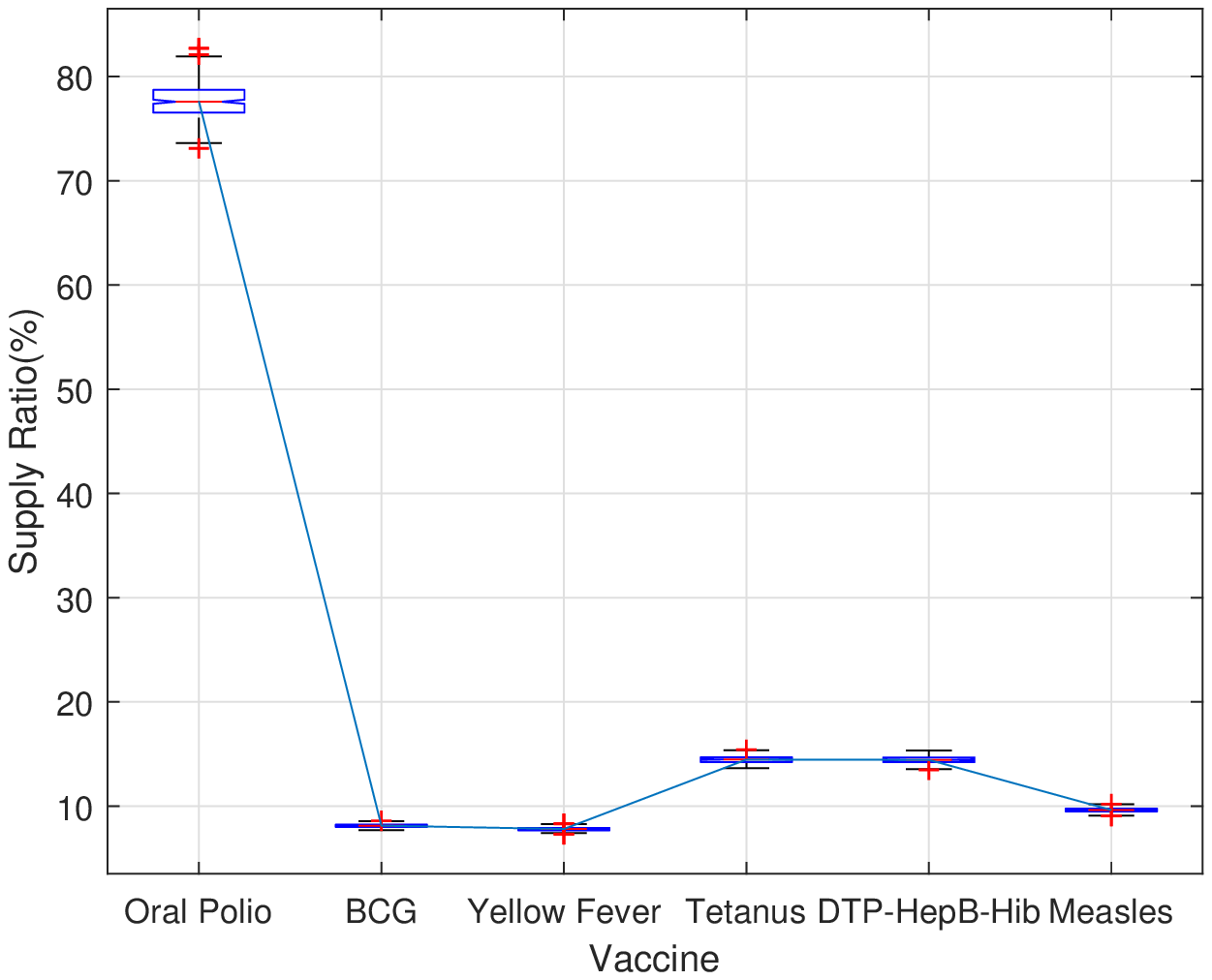}
  	\caption{Four-tier Supply Chain}
	\label{fig:SR_Base_4Tiers}
    \end{subfigure}

    \begin{subfigure}{0.5\textwidth}
        \centering
 	   	\includegraphics[width=\textwidth]{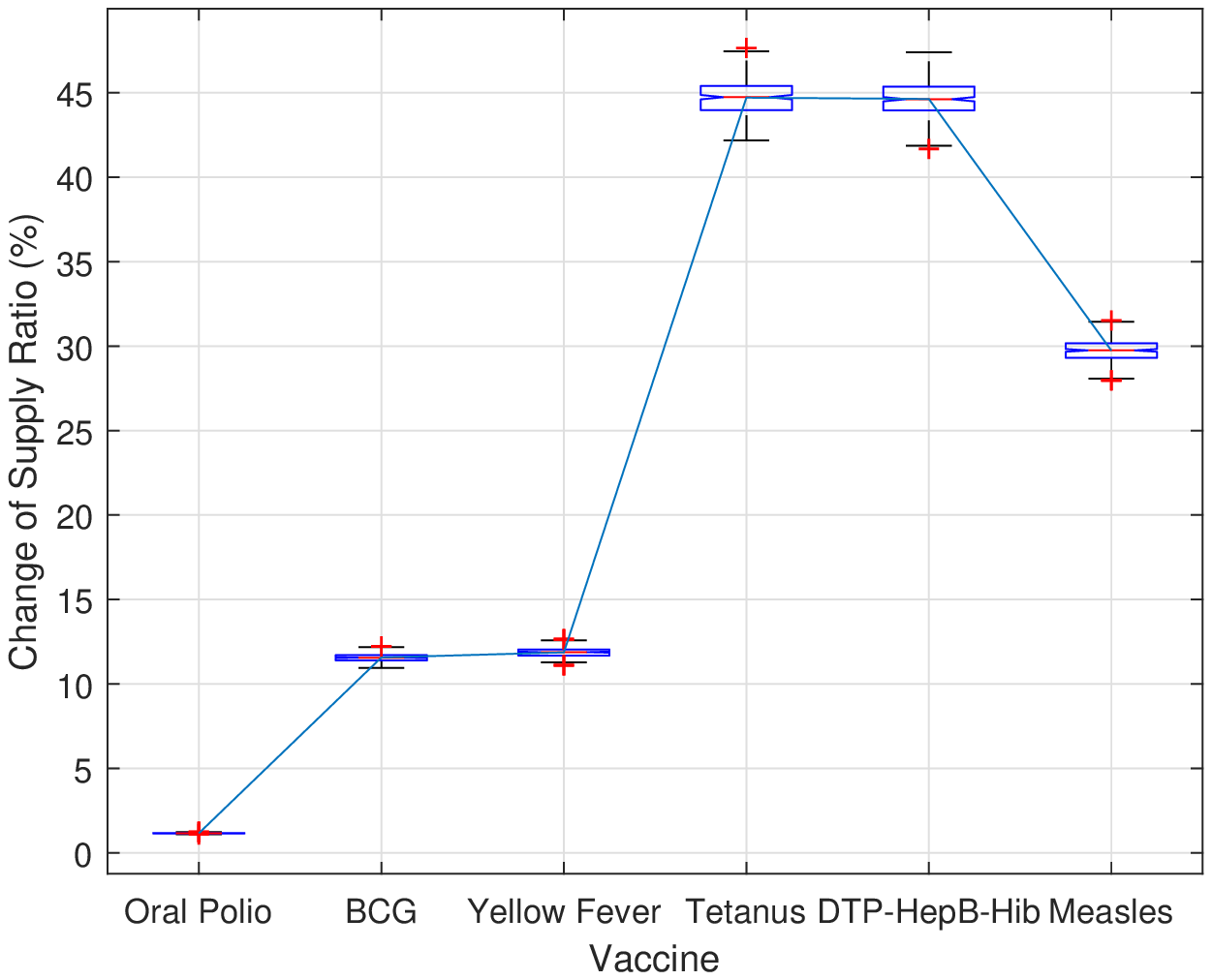}
   	\caption{Capacity Added to Clinics}
	\label{fig:SR_Base_3TiersIHC}
    \end{subfigure}%
    \begin{subfigure}{0.5\textwidth}
        \centering
 	   	\includegraphics[width=\textwidth]{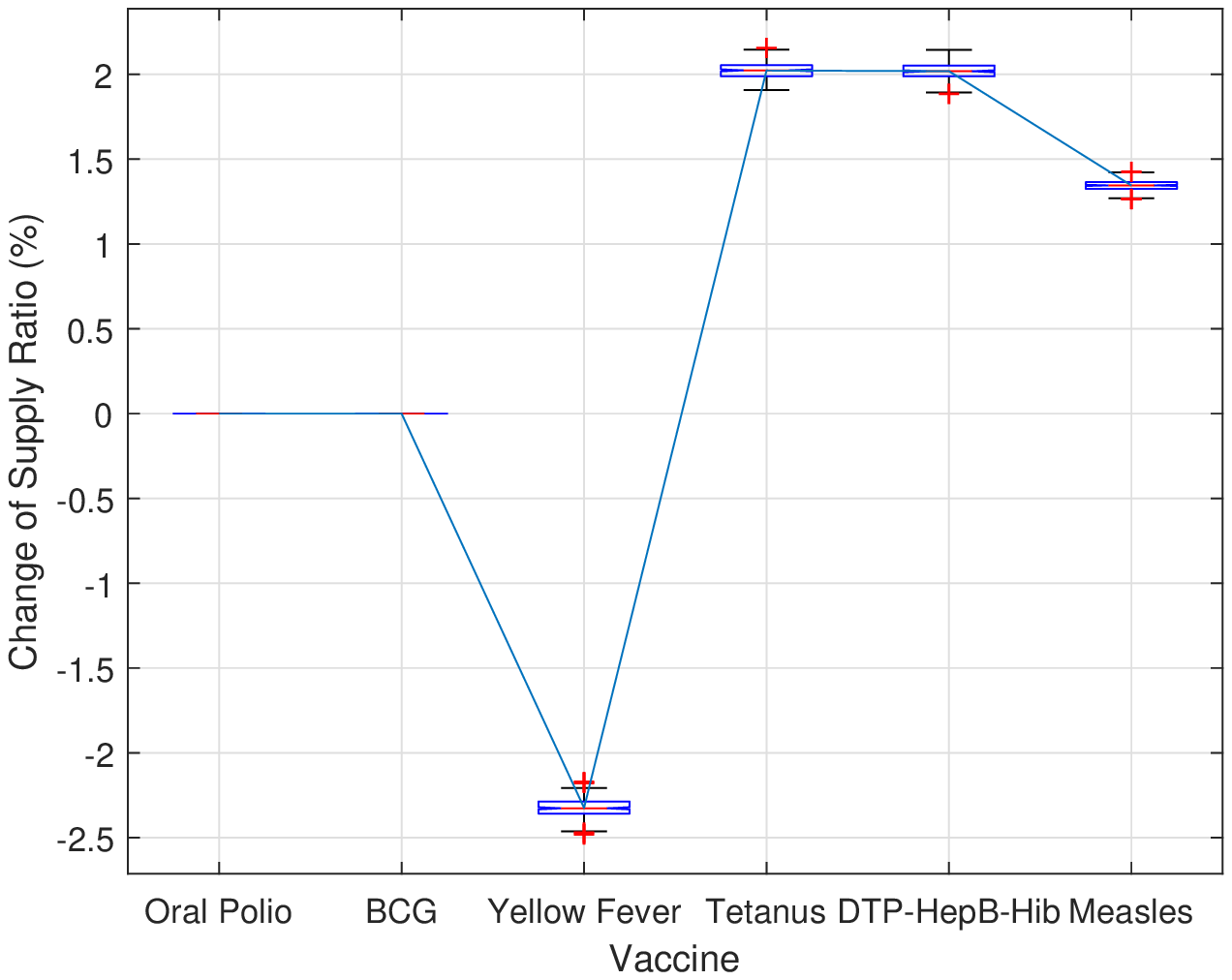}
  	\caption{Capacity Added to  District Stores}
	\label{fig:SR_Base_3TiersDist}
    \end{subfigure}
    \caption{(a) Vaccine-Based SR for Four-Tier Supply Chain; (b)  \% Increase in SR in a Three-Tier Supply Chain if capacity is added to Clinics;  (c) \% Increase in SR in a Three-Tier Supply Chain if capacity is added to District Stores.} \label{fig::SRRegions_4and3}
\end{figure}

\begin{figure}[h] 
\centering
    \begin{subfigure}{0.5\textwidth}
        \centering
 	   	\includegraphics[width=\textwidth]{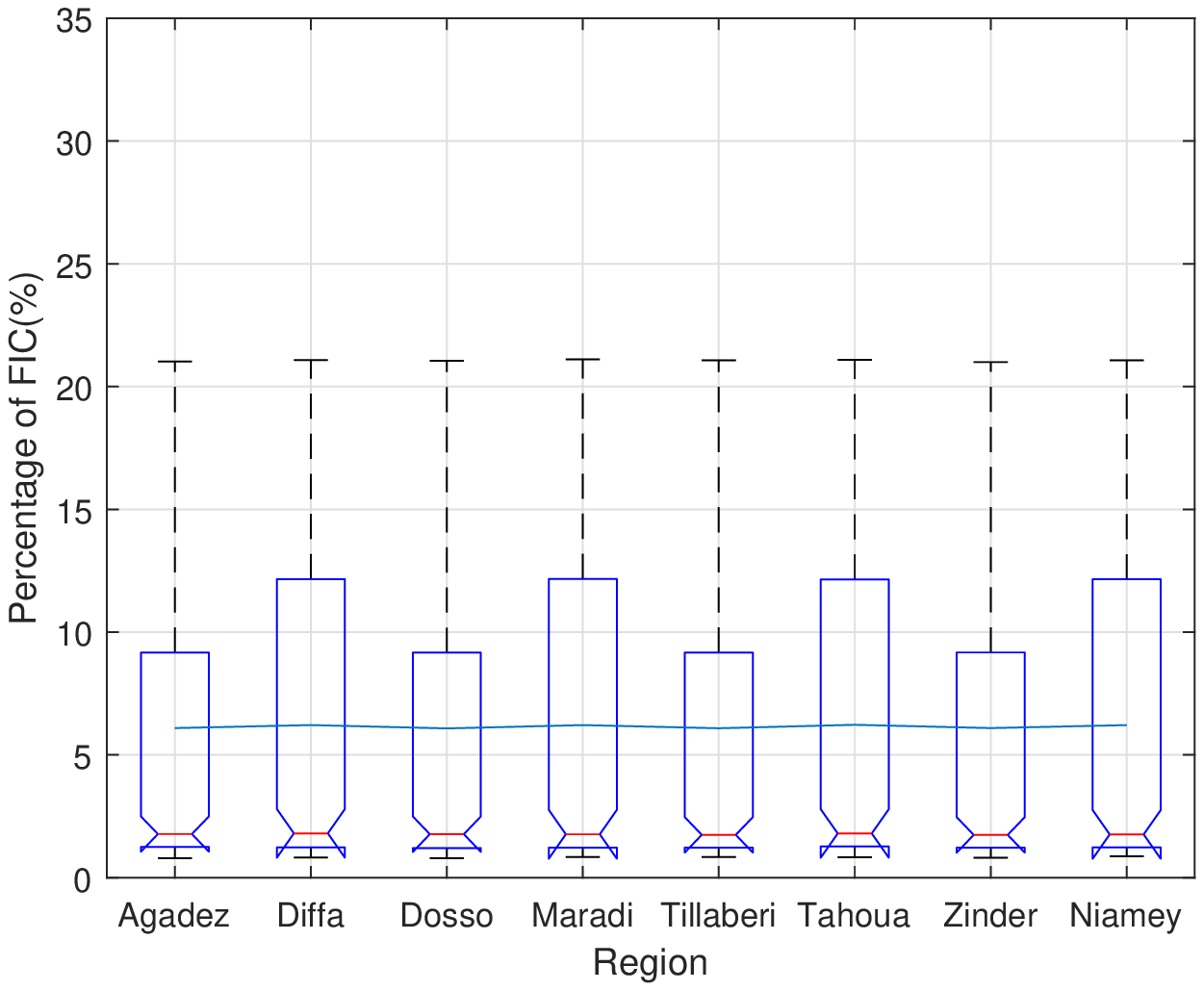}
   	\caption{FIC}
	\label{fig:FIC_Mixed_M}
    \end{subfigure}%
    \begin{subfigure}{0.5\textwidth}
        \centering
 	    		\includegraphics[width=\textwidth]{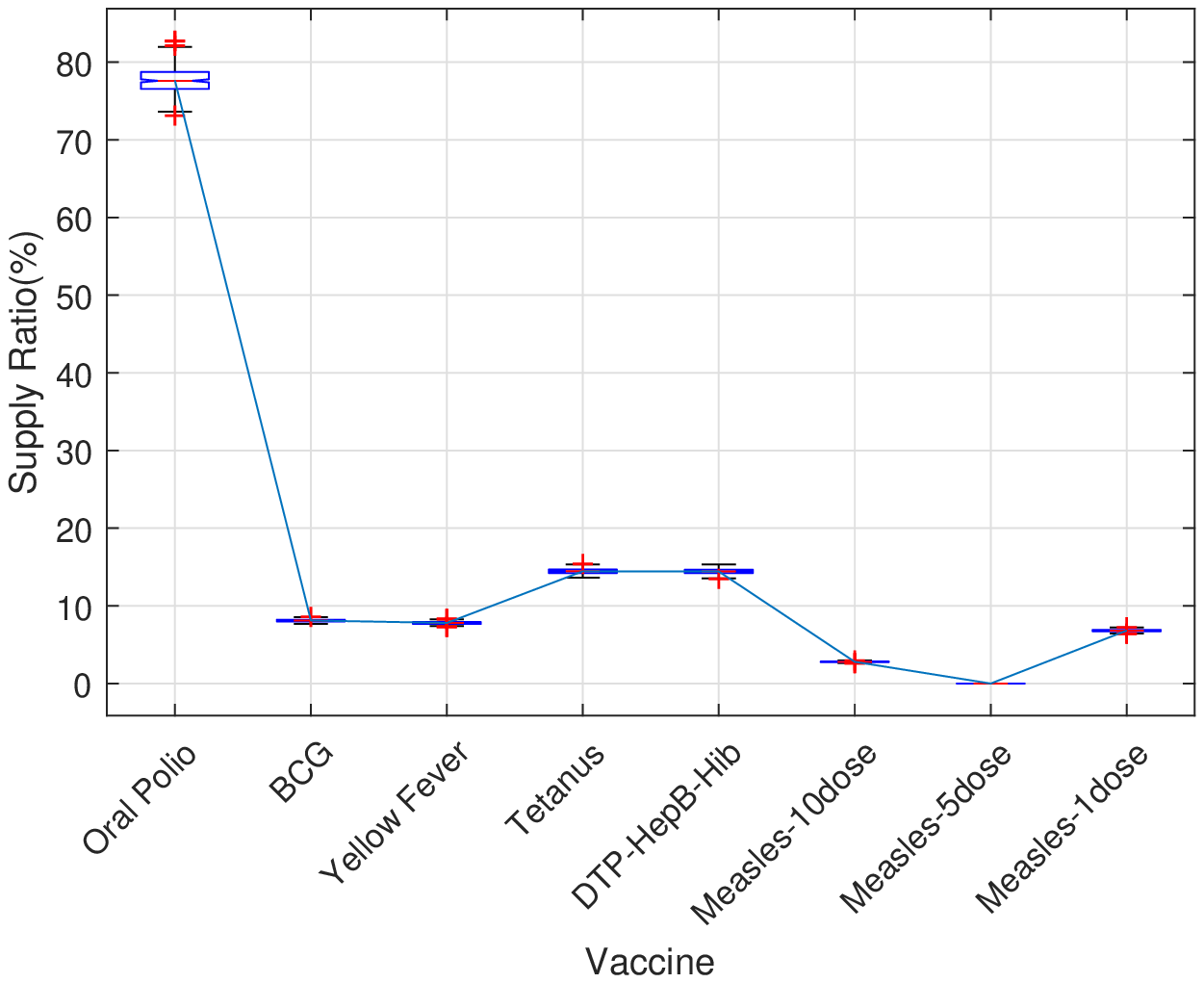}
  	\caption{SR}
	\label{fig:SR_Mixed_M}
    \end{subfigure}
    \caption{Scenario 1: region-based FIC and vaccine-based SR.} \label{fig::Mixed_M}
\end{figure}

\begin{figure}[h] 
        \centering
 	 \includegraphics[width=0.5\textwidth]{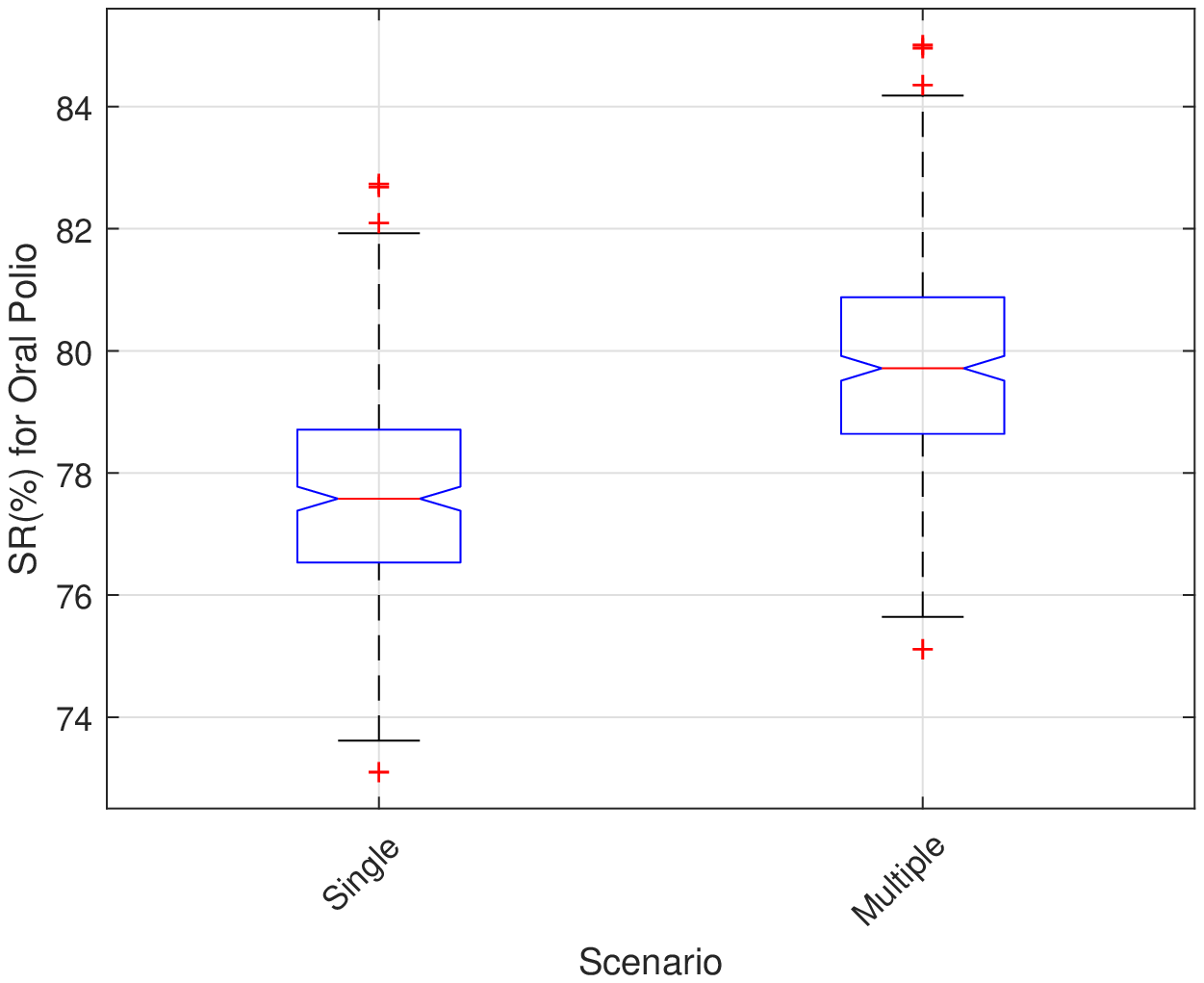}
	    \caption{Scenario 2: SR of oral polio vaccine for single and multiple vial sizes.} \label{fig:SROPV_Mixed_MB}
\end{figure}

\begin{figure}[h] 
\centering
    \begin{subfigure}{0.5\textwidth}
        \centering
 	   	\includegraphics[width=\textwidth]{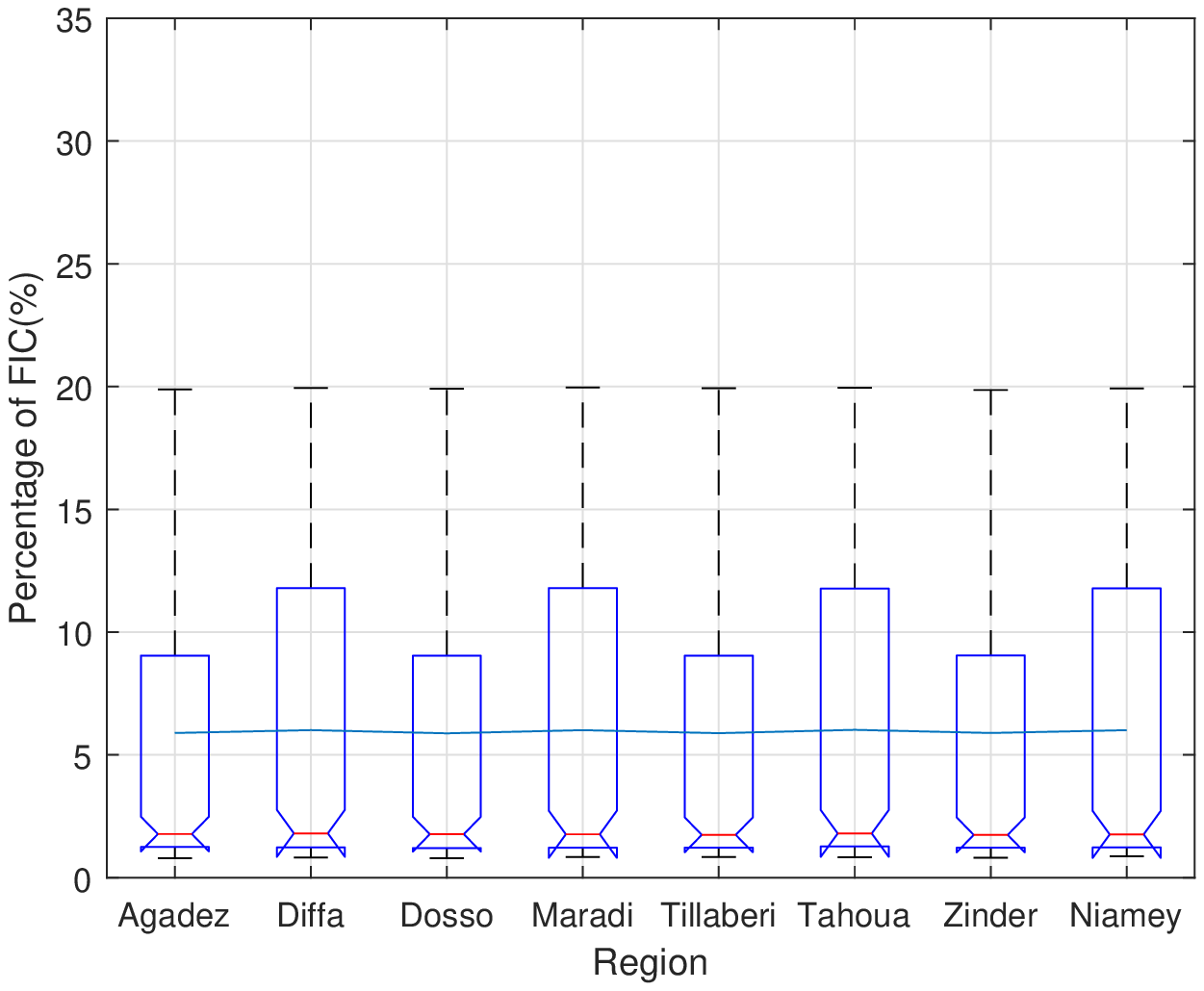}
   	\caption{FIC}
	\label{fig:FIC_Base_DualCham}
    \end{subfigure}%
    \begin{subfigure}{0.5\textwidth}
        \centering
 	    		\includegraphics[width=\textwidth]{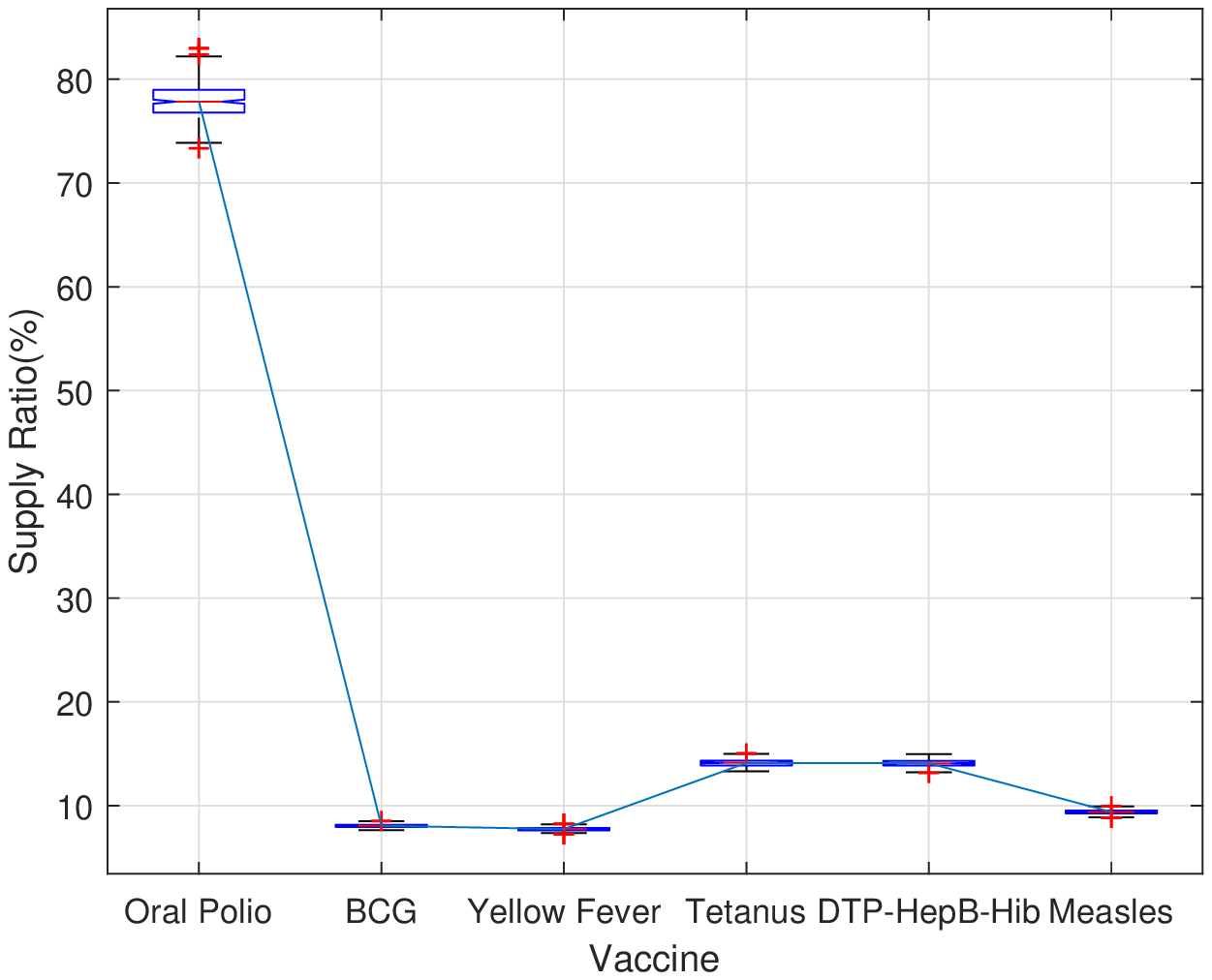}
  	\caption{SR}
	\label{fig:SR_Base_DualCham}
    \end{subfigure}
    \caption{Region-based FIC and vaccine-based SR for a four-tier supply chain with dual-chamber measles vaccine.} \label{fig::DualCham}
\end{figure}

\begin{figure} [h] 
        \centering
 	    \includegraphics[width=0.5\textwidth]{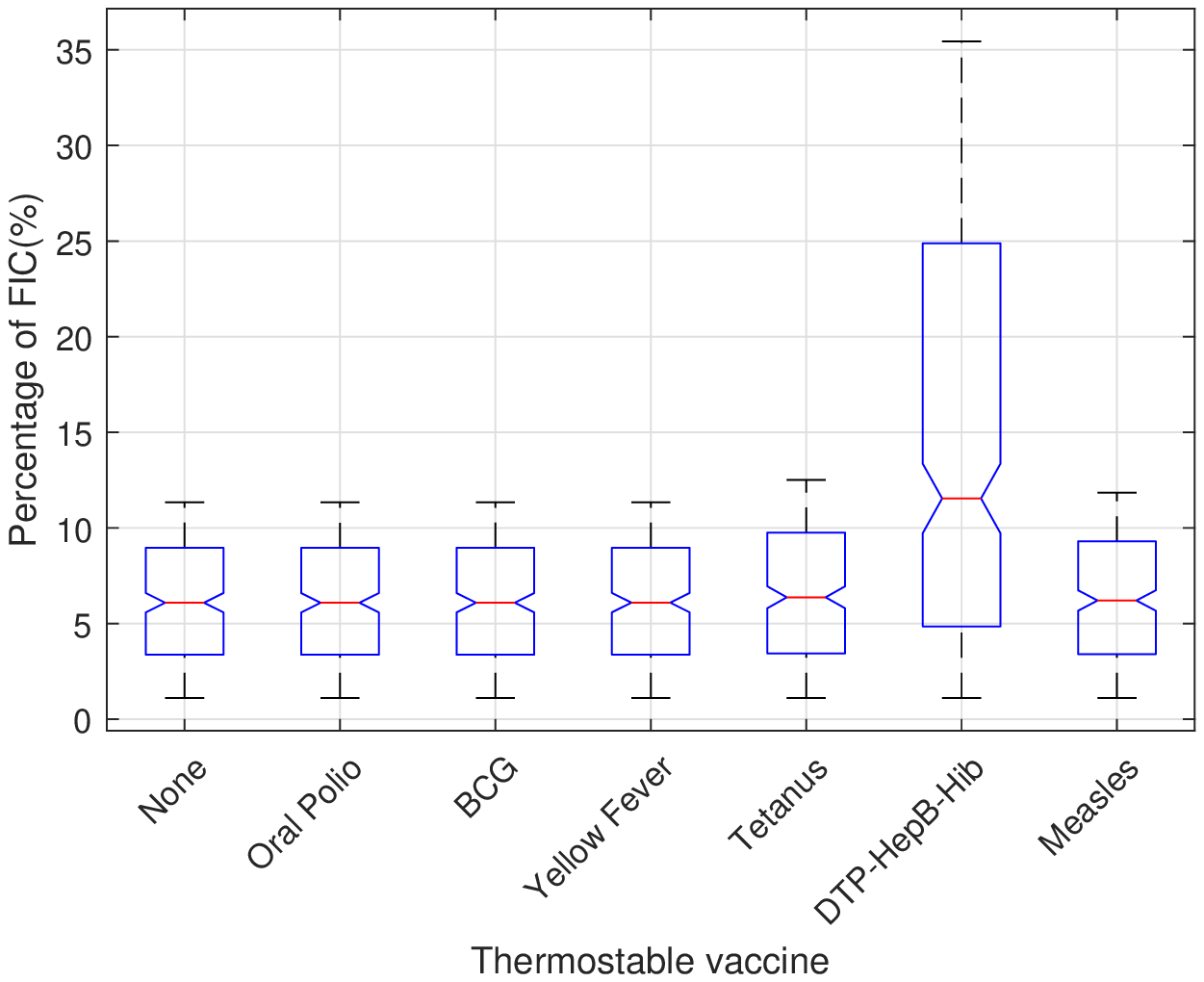}
  	\caption{Country-Based FIC if the labeled Vaccine Is Thermostable.}
	\label{fig:FIC_Ther}
\end{figure}

\begin{figure} 
\centering
    \begin{subfigure}{0.5\textwidth}
        \centering
 	    \includegraphics[width=\textwidth]{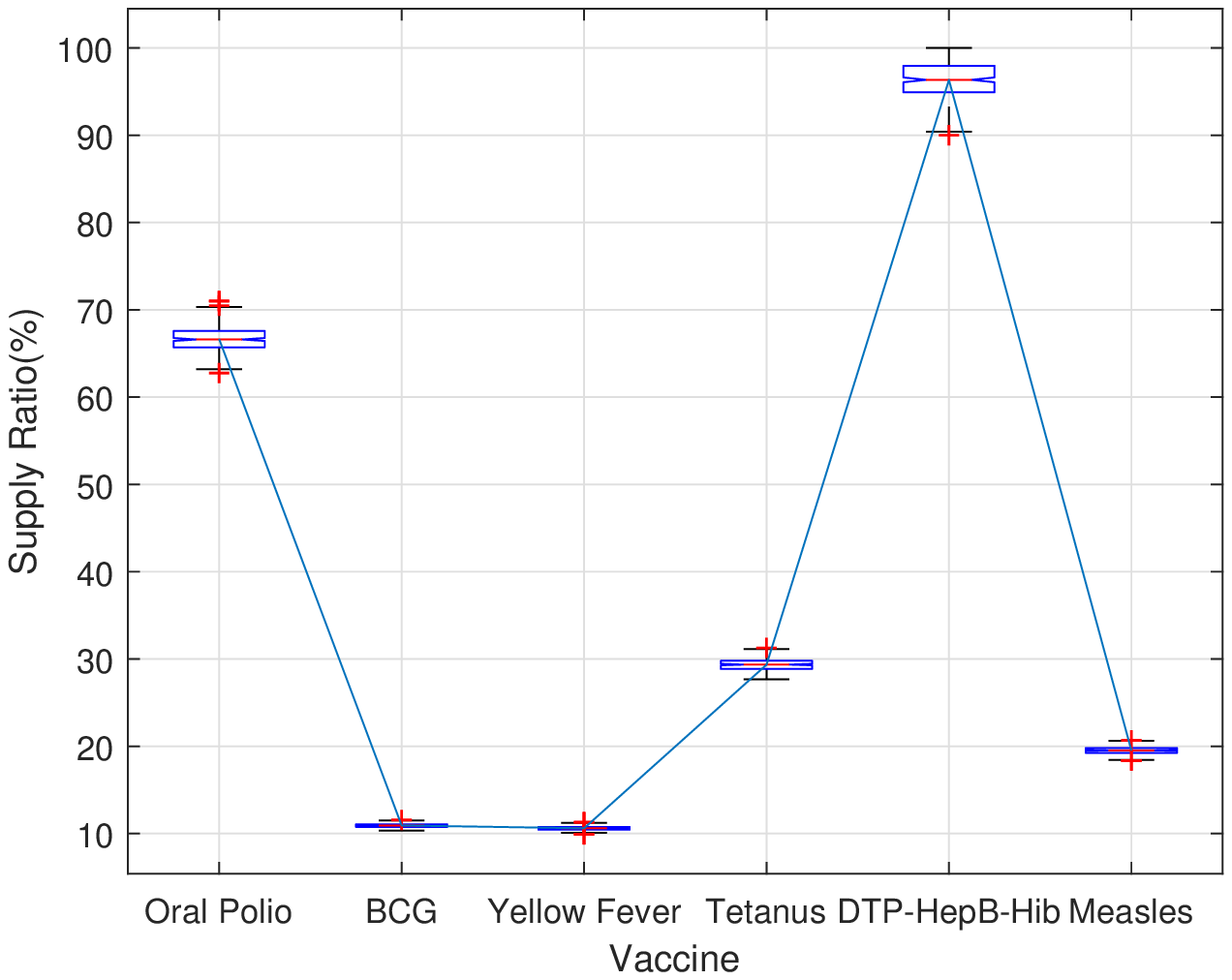}
  	\caption{DTP-HepB-Hib}
	\label{fig:SR_Ther_DTP}
    \end{subfigure}%
    \begin{subfigure}{0.5\textwidth}
        \centering
 	    	\includegraphics[width=\textwidth]{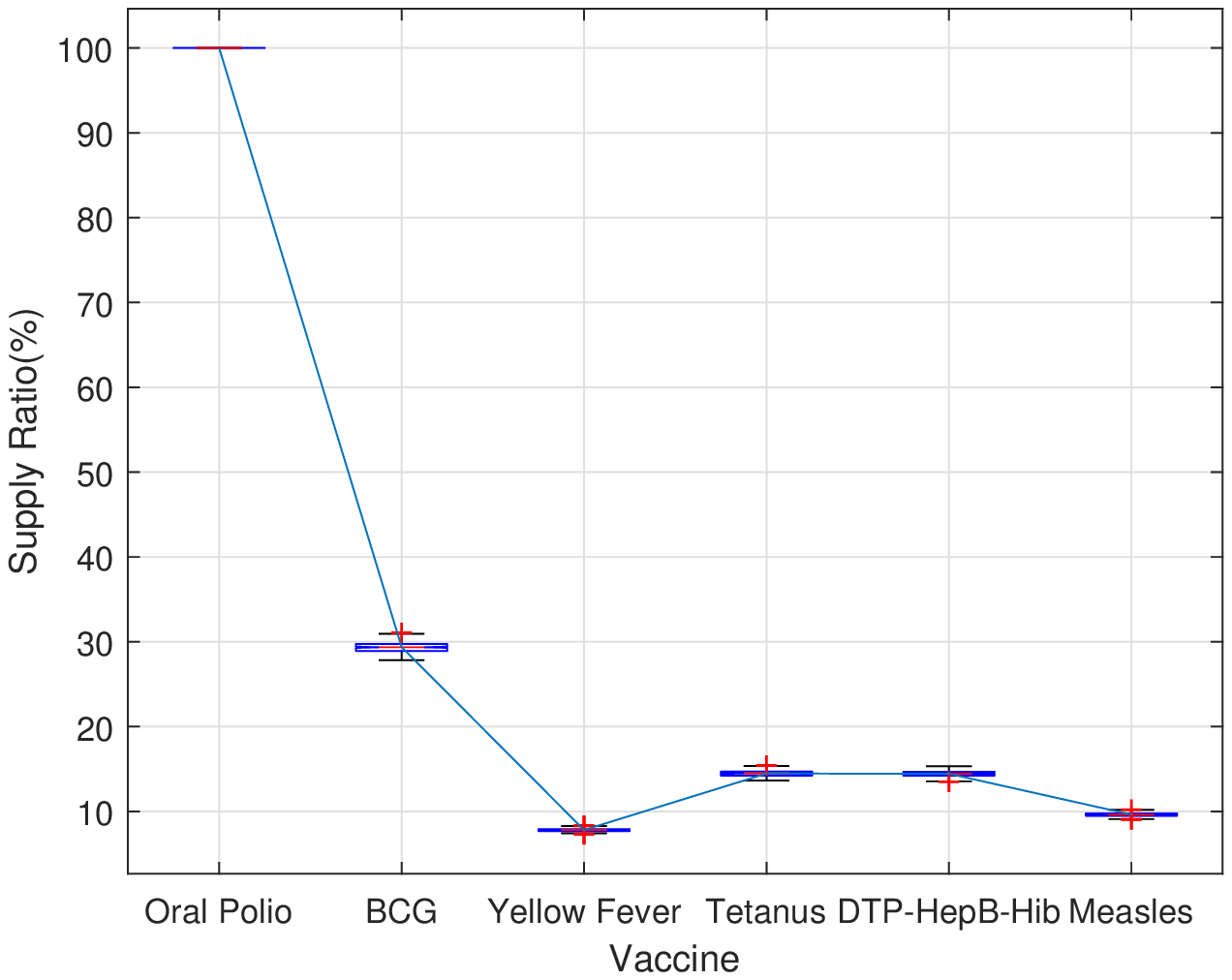}
  	\caption{polio}
	\label{fig:SR_Ther_OPV}
    \end{subfigure}
    \begin{subfigure}{0.5\textwidth}
        \centering
 	    \includegraphics[width=\textwidth]{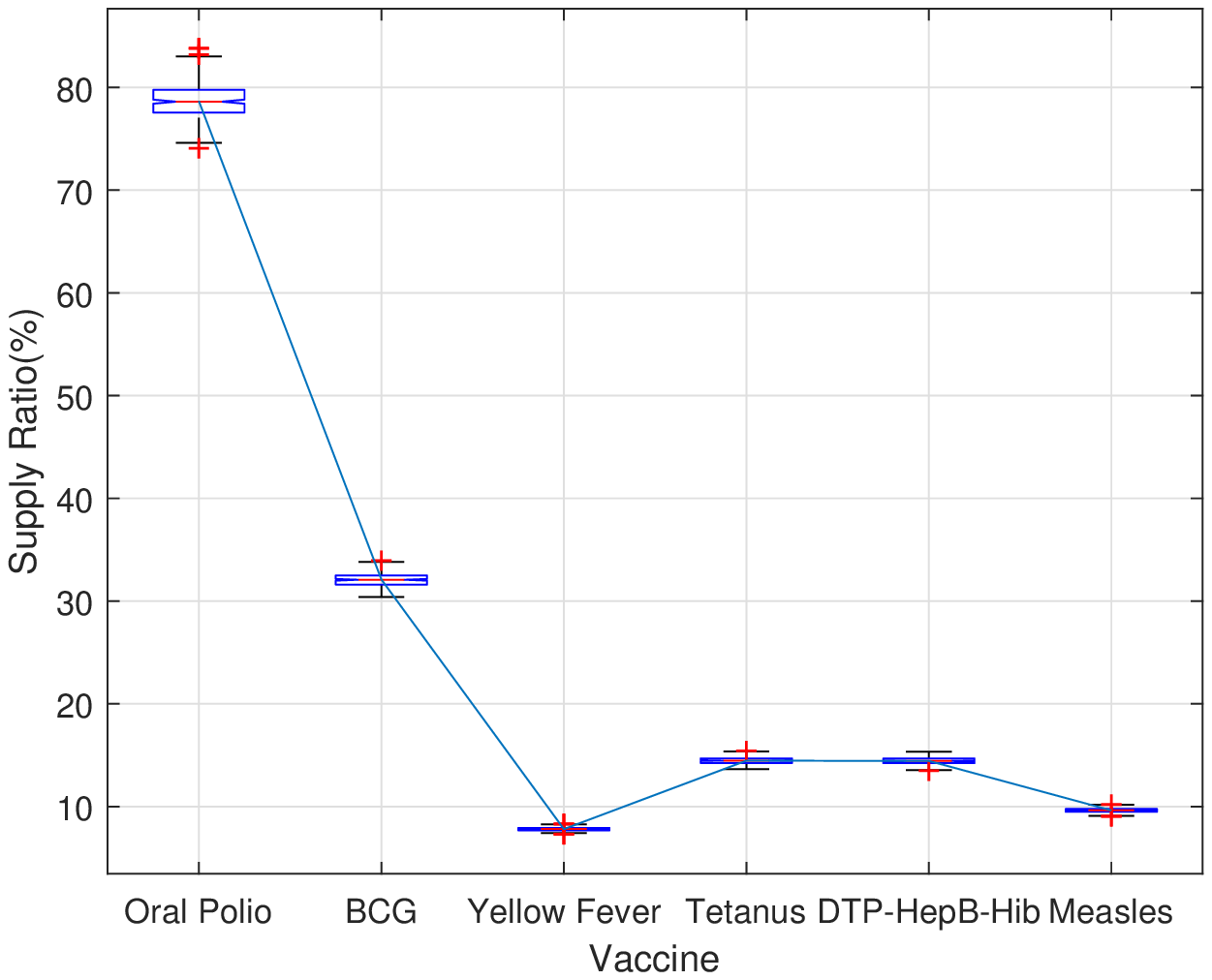}
  	\caption{BCG}
	\label{fig:SR_Ther_BCG}
    \end{subfigure}%
    \begin{subfigure}{0.5\textwidth}
        \centering
 	    	\includegraphics[width=\textwidth]{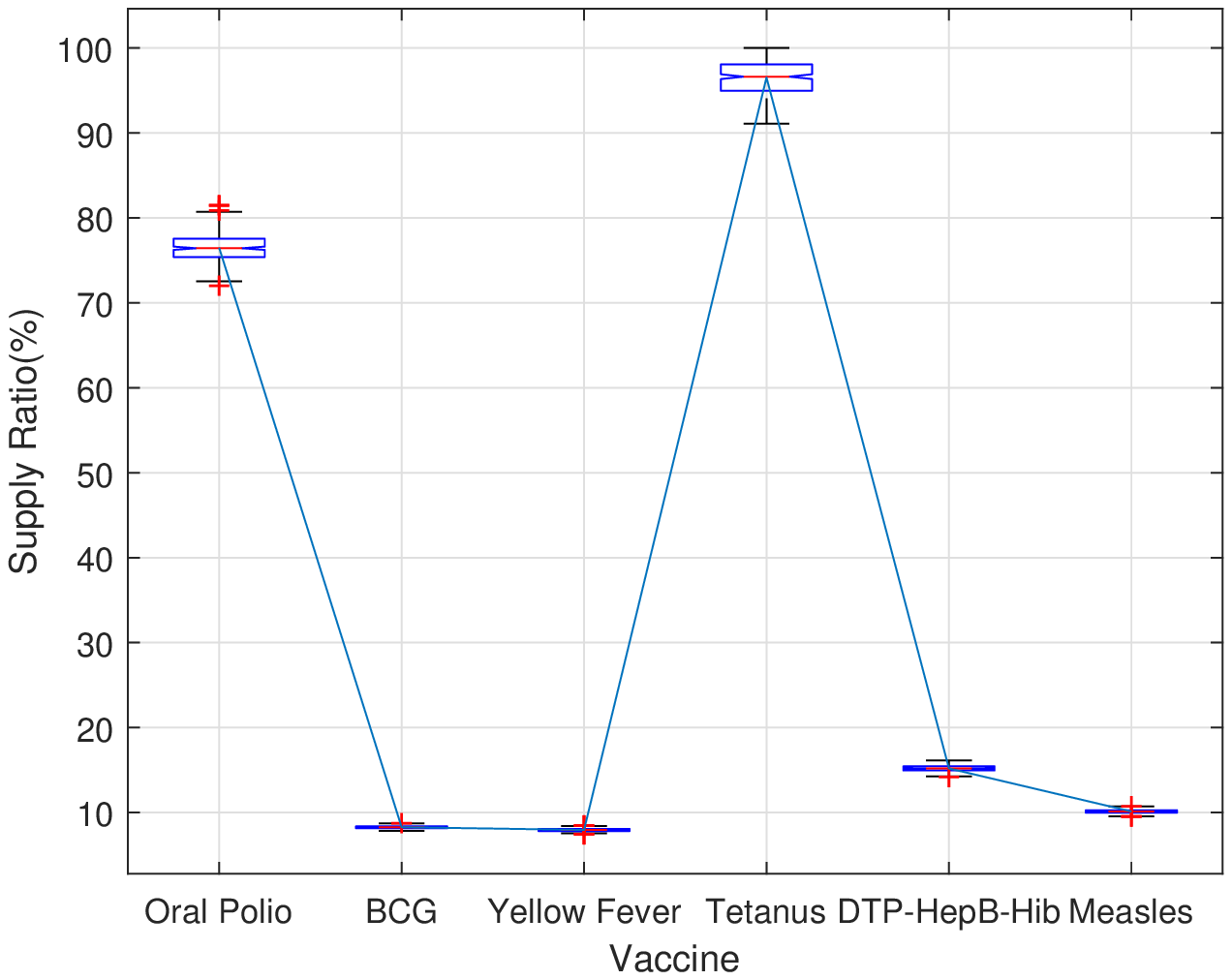}
  	\caption{Tetanus}
	\label{fig:SR_Ther_Tetanus}
    \end{subfigure}

        \begin{subfigure}{0.5\textwidth}
        \centering
 	    \includegraphics[width=\textwidth]{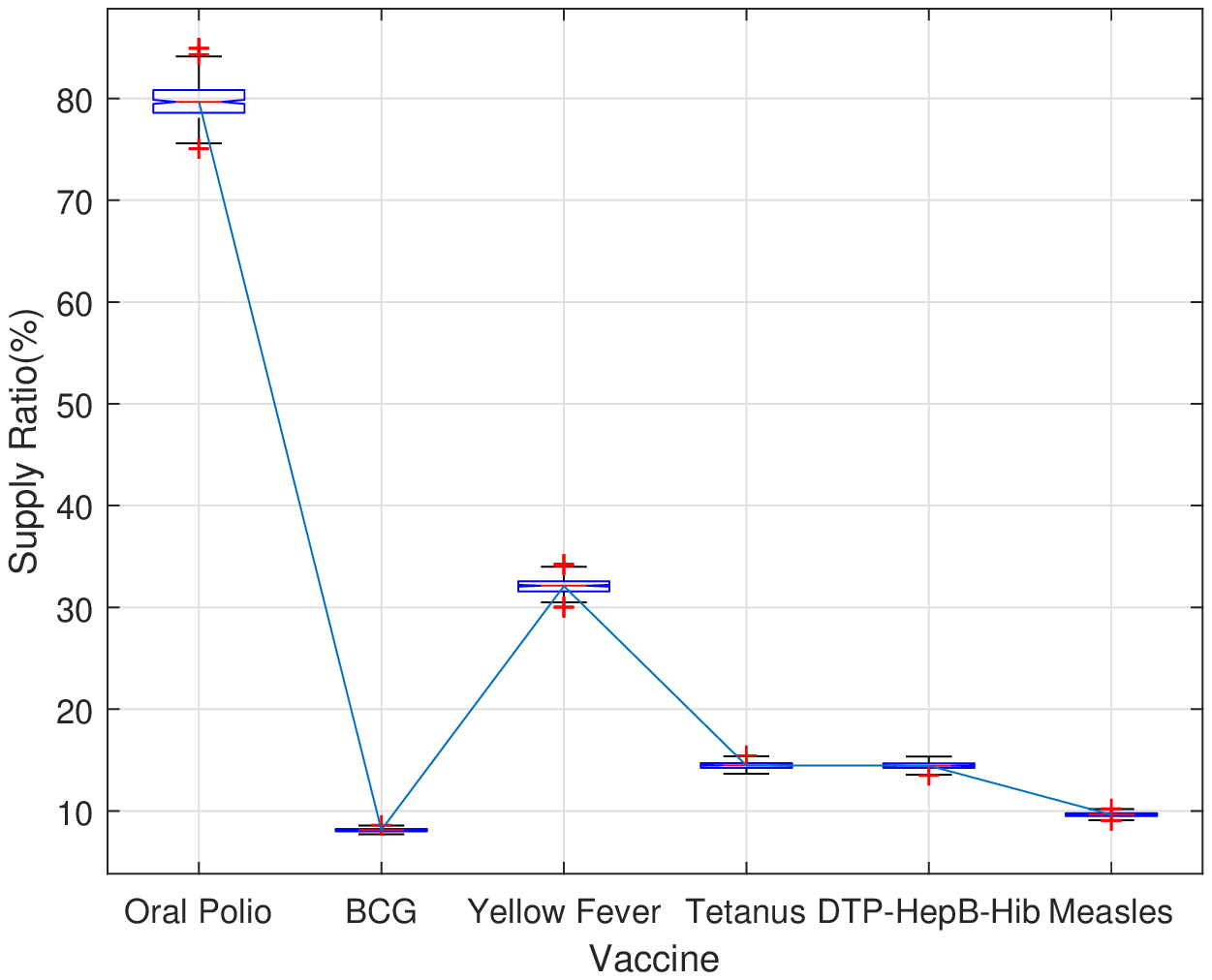}
  	\caption{YF}
	\label{fig:SR_Ther_YF}
    \end{subfigure}%
    \begin{subfigure}{0.5\textwidth}
        \centering
 	    	\includegraphics[width=\textwidth]{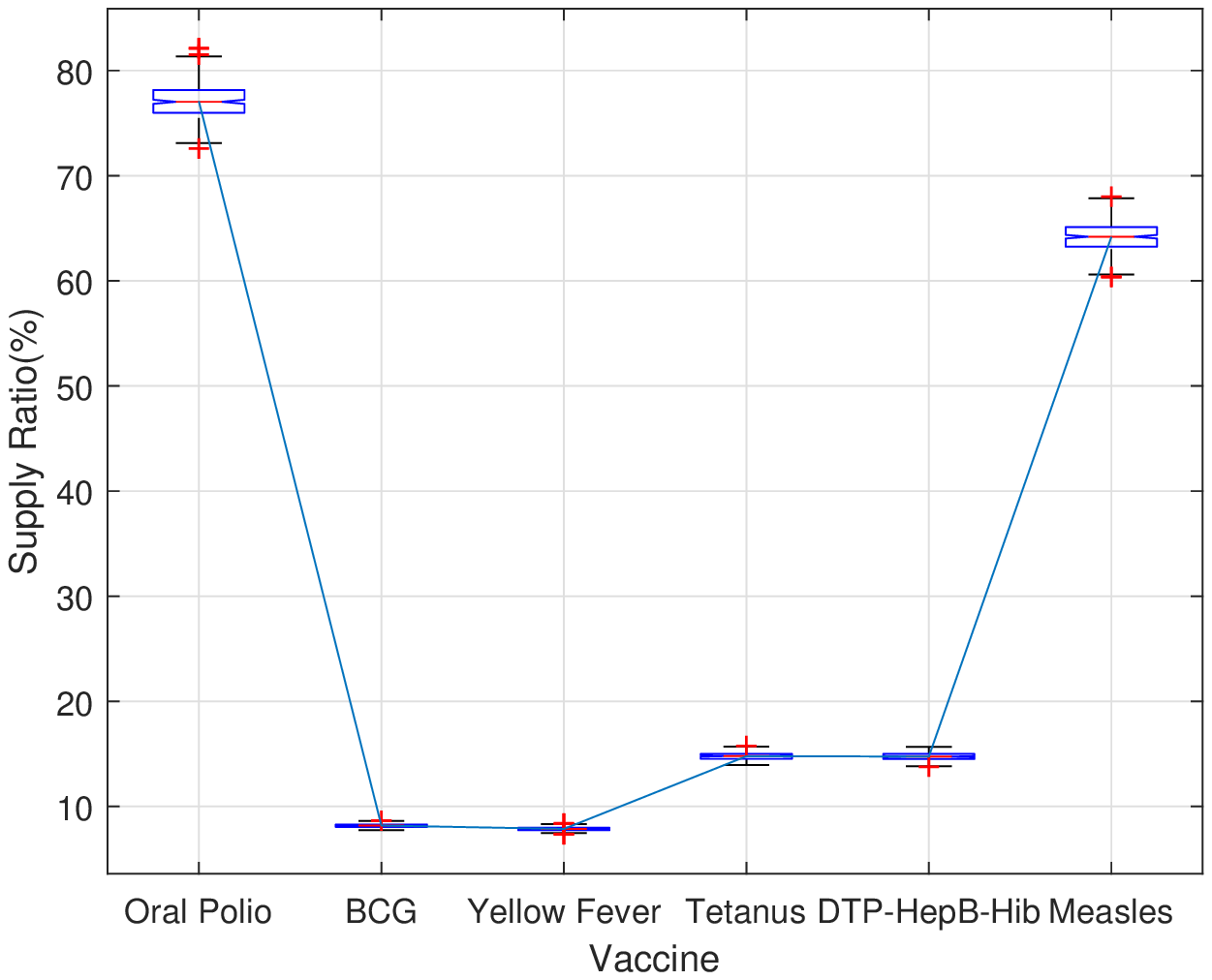}
  	\caption{Measles}
	\label{fig:SR_Ther_Measles}
    \end{subfigure}

    \caption{Vaccine-based SR: the vaccine in a panel's caption is thermostable.} \label{fig::Thermos}
\end{figure}

Numerical analysis offers answers to the research questions outlined in the introduction section. Our experimental results are summarized in Figures \ref{fig::FICRegions_4and3} to \ref{fig::Thermos}. These figures present the distribution of FIC and SR via box and whisker plots. The lines in these plots represent the mean values.

\medskip

\noindent{\bf R1: How does changing the number of tiers in the supply chain affect vaccine availability?}
We evaluate the impacts on FIC and SR of removing regional stores in Niger's current four-tier supply chain. 
In the three-tier supply chain, the district stores receive shipments from the central store and deliver vaccines to clinics.

We test two scenarios: one in which the cold storage capacity of regional stores is relocated to the district stores, and one in which the cold storage capacity is equally distributed among clinics. Next, we summarize our  observations.


The four-tier supply chain has an average FIC below 10\% and a maximum FIC below $22\%$ (see \figref{fig:FIC_Base_4Tiers}). Thus, immunization coverage rates are far from the target value of 90\% established by the World Health Assembly.  Removal of regional stores leads to higher values of average FIC and SR, assuming the cold storage capacity is reallocated downstream in the supply chain.

The values of SR and FIC are highest in the three-tier supply chain when cold storage capacity is relocated to clinics. Notice that the distribution of FIC is right-skewed 
(see \figref{fig:FIC_Base_3TiersIHC}). The mean value of FIC and the corresponding right tail is shorter when the capacity is relocated to the district stores  (see \figref{fig:FIC_Base_3TiersDist}).

This observation goes against
the risk-pooling concept in inventory management, 
but it is consistent with the work of \cite{Kurata14}, who studies a supply chain with a single supplier, multiple retailers and product-availability-conscious customers (who abandon a given purchase if the items are not available at a given store). \cite{Kurata14} shows that in such a system locating inventory at retailers outperforms locating inventory at the supplier. 
In the environment of our case study, 
if a vaccine is not available, a child is not vaccinated. Thus, our patients are product-availability-conscious.  Furthermore, demand across the supply chain is highly correlated, so risk-pooling should be less effective.

In the four-tier supply chain, the average FIC across all regions is $9.5\%$, and the average SR across all vaccine types is $30\%$. A paired t-test at significance level 0.05 indicates that the mean values of both FIC and SR  are statistically better in three-tier supply chains than they are in four-tier supply chains. This observation contradicts the results presented in \cite{chen2014planning}.

The largest SR is achieved for oral polio vaccine in  four-tier supply chains (\figref{fig::SRRegions_4and3}a).  Thus, it has little room for improvement (\figref{fig::SRRegions_4and3}b) when the supply chain is reduced to three tiers.  The reasons for this are that this vaccine ($i$)  requires four doses to complete the vaccine regimen, so demand for this vaccine is higher than others; ($ii$) takes less storage space; and ($iii$) uses a diluent that does not need refrigeration.

\medskip
\noindent{\bf R2: How does vaccine vial size effect immunization coverage rates?} 
We evaluate the  impact of  manufacturing vaccines in different vial sizes (i.e., 1-, 5-, and 10-dose vials) on FIC and SR in Niger by extending  model~(P) 
to consider a combination of vaccine vial sizes. The corresponding mathematical model is presented in Appendix \ref{appC}.

We consider two scenarios. The first assumes that only the measles vaccine is distributed in vials of different sizes (1-, 5-, and 10-doses), and the second assumes that both measles and BCG vaccines are distributed in vials of different sizes (measles: 1-, 5-, and 10-doses;  BCG:  10- and 20-doses). In each scenario, the vial sizes for the remaining vaccine types are as presented in Table \ref{tab-VaccineChar}.

Using a combination of different sized vials does not lead to statistically significant improvements of FIC in LMI countries with limited cold supply chain capacity (compare  Figures~\ref{fig:FIC_Base_4Tiers} and~\ref{fig:FIC_Mixed_M}). The percentages of FIC in different regions do not change because, even though using a combination of vials of different sizes leads to lower OVW and improved vaccine availability \citep{azadi2019developing}, using a combination of sizes requires additional refrigeration space. The results from both scenarios support this observation.

The improvements in SR when only the measles vaccine is distributed in vials of different sizes are not statistically significant (see  Figures~\ref{fig:SR_Base_4Tiers} and~\ref{fig:SR_Mixed_M}). However, there is a statistically significant improvement in SR when both measles and BCG vaccines are distributed in vials of different sizes (Scenario 2).  \figref{fig:SROPV_Mixed_MB} indicates a 2\% increase in the  mean value of SR for the polio vaccine when  combinations of sizes are used for both measles and BCG. A paired t-test at significance level $0.05$  shows that this difference is statistically significant. A $2\%$ increase in  vaccine availability can potentially immunize an additional $20,000$  children.

\medskip
\noindent {\bf R3: How does  using new vaccine presentations affect immunization coverage rates?} 
To evaluate the impacts of using dual-chamber and thermostable vaccines on FIC and SR in Niger, consider two scenarios. In the first,  replace regular measles vaccines with dual-chamber injection devices. In the second, replace one of the EPI vaccines with a thermostable vaccine, which requires no cold storage space.

The difference in the mean distribution of FIC and SR for measles between using a single-dose dual-chamber injection device and using  ten-dose vials of measles vaccine (the current implementation) is statistically insignificant (compare  \figref{fig::DualCham}  to \figref{fig:FIC_Base_4Tiers}
and \figref{fig:SR_Base_4Tiers}). This finding also holds when comparing a single-dose dual-chamber injection device to a single-dose standard vial.  These results are confirmed via a paired $t$-tests at a significance level of 0.05.

Using a dual-chamber vaccine consumes more of the limited storage space, 
but it also reduces OVW. Thus, the compound effect led to a slight reduction in FIC and SR.

The country-based FIC of Niger could almost double if the classically available DTP-HepB-Hib vaccines are substituted with thermostable vaccines (\figref{fig:FIC_Ther} shows that the average FIC is 6\% if no thermostable vaccine is used). Each vial of DTP-HepB-Hib vaccine contains only one dose, and its packed volume is the largest among all vaccine types, so using thermostable vaccines would free much cold storage space that can be used for other vaccines.

Comparing \figref{fig:SR_Base_4Tiers} and \figref{fig::Thermos} yields the following conclusions about the effects of using a thermostable formulation for a single vaccine.
\begin{itemize}
\item
Using a thermostable DTP-HepB-Hib vaccine increases the mean value of SR for every vaccine but polio (\figref{fig:SR_Ther_DTP}). This small decrease of polio's SR arose from the increase in the total number of children vaccinated, which is the denominator $\phi_j^s$ in the definition of SR (Equation~\eqref{eq-FICRatio}), and from polio's high original SR.
\item
Using a thermostable oral polio vaccine increases the corresponding mean value of its SR to 100\%. Additionally, the mean value of SR for BCG increases by close to three times (\figref{fig:SR_Ther_OPV}). The SR values for the remaining vaccines do not change.
\item
Using a thermostable tetanus vaccine increases the corresponding mean value of its SR by close to four times. Additionally, this increases the mean values of SR for BCG and YF by $4\%$, and the mean values for both DTP-HepB-Hib and measles by $8\%$ (\figref{fig:SR_Ther_Tetanus}).
\item
Using a thermostable YF vaccine increases the corresponding mean value of its SR by close to four times (\figref{fig:SR_Ther_YF}). The SR values for the remaining vaccines do not change.
\item
Using a thermostable measles vaccine increases the mean of SR for every vaccine except oral polio (\figref{fig:SR_Ther_Measles}).
\end{itemize}

\medskip
\noindent {\bf Recommendations:}  These observations lead to the following recommendations:

We recommend that public health authorities invest to increase cold storage capacity (i.e., cold rooms, refrigerators, and freezers) of clinics because this is the bottleneck in this supply chain, so these investments would directly lead to improved immunization coverage. 
Countries with limited resources should prioritize cold storage and distribution for those vaccines with the highest number of doses required by vaccine regimen (i.e., polio) because the availability of these vaccines has the greatest influence on SR.  Niger exemplifies the benefit of this practice by having a very high SR for the oral polio vaccine.

We recommend that public health authorities  evaluate the trade-offs among OVW, FIC, and SR when considering changes to vaccine packaging. Our numerical analysis indicates that the use of a combination of vials of different sizes of the same vaccine leads to reduced OVW, but may not necessarily lead to more FIC and higher SR for every vaccine type in LMI countries with limited storage capacity in the cold supply chain.

We recommend that public health authorities evaluate the trade-offs among vaccine presentation, storage space required, vaccine availability, and  waste from OVW when they consider using new vaccine technologies, such as dual-chamber and thermostable vaccines. This is especially relevant in countries that have limited resources because new technologies are expensive and may require additional storage space.

We do not recommend the
use of dual-chamber vaccines. They require additional cold storage space per dose because the diluent in a dual-chamber vaccine is attached to the vaccine (the diluent of a regular  vaccine does not need refrigeration). Hence, their use reduces the availability of all vaccines. 
These problems greatly outweigh dual-chamber vaccines'  zero OVW and ease of administration.

Thermostable vaccines reduce waste by consuming less cold storage space.  Thus, they increase immunization coverage. However, thermostable vaccines are expensive, so public health authorities in LMI countries should use their limited resources strategically by investing in thermostable vaccines that have the greatest impact on immunization coverage. Therefore, we recommend vaccines that require large storage capacity per dose, such as DTP-Hep-Hib, have thermostable formulations.

\section{Conclusions and Future Research} \label{sec-Conclusion}
Public health officials in low- and middle-income (LMI) countries have historically used educational campaigns to teach the importance of vaccination and increase participation.  These campaigns are a good first step, but their success depends on the efficiency of the vaccine supply chain to deliver  EPI (Expanded Program on Immunization of the World Health Organization) vaccines to clinics. Thus, public health officials need tools that evaluate their decisions' impacts on the performance of the supply chain. Such evaluations could guide public policies that improve immunization coverage rates.

We develop a stochastic optimization model that captures the uncertainty of demand for vaccination via chance constraints.  To the best of our knowledge, this study is the first to apply a stochastic optimization model to evaluate  the impact of supply chain decisions on LMI countries' immunization coverage rates, as measured by the proportion of fully immunized children (FIC) and the supply ratio (SR) for a case study using data from Niger.  The specific supply chain issues examined are the supply chain design, the vaccine packaging, and the vaccine presentation. 

We conduct a comprehensive  analysis of the effects that the following supply chain decisions have on immunization coverage rates: ($i$) redesigning the supply chain from a four- to a three-tier supply chain by removing the regional stores; ($ii$) distributing a combination of vaccine vials with different sizes; and ($iii$) using new technologies, such as dual-chamber injection devices and thermostable vaccines. Our main observations are

\begin{enumerate}
\item Streamlining the supply chain by removing regional stores leads to higher values of average FIC and SR if the cold storage capacity is reallocated to the clinics.

\item In LMI countries with limited cold supply chain resources, using a combination of vials of different sizes for one vaccine leads to reduced open vial wastage (OVW) but may not necessarily lead to increased FIC and SR. 

\item Public health officials need to understand the trade-offs between vaccine availability and the waste from OVW when considering the use of dual-chamber vaccines.  This is especially relevant in countries that have limited resources because  dual-chamber vaccines require additional storage space per vial, which reduces the availability of all vaccines.

\item Using thermostable vaccines reduces waste and therefore increases immunization coverage. Thermostable vaccines are expensive, but public health authorities in LMI countries can use their limited resources strategically by investing in this technology for those vaccines for which doing so will cause the greatest increase in immunization coverage. For example, DTP-Hep-Hib (diphtheria, tetanus, pertussis, hepatitis B and haemophilus influenza) requires large storage capacity per dose, so  substituting DTP-Hep-Hib with a thermostable vaccine would reduce waste and release the cold storage space to other vaccines.
\end{enumerate}

As with other studies of humanitarian supply chains  that focus on  minimizing the delivery time of essential supplies, minimizing stock-outs, maximizing the number of people served, etc.,  this study does not explicitly focus on costs.  Reductions of supply chain costs, which are funded by non-profit organizations (World Health Organization, World Bank, United Nations Children's Fund, etc.) and the local government, clearly would lead to savings that could be used to extend cold storage capacity. Thus, we plan to extend this analysis to consider the economic impacts of the supply chain decisions discussed in this paper.



\bibliographystyle{ormsv080} 
\bibliography{ref} 

\begin{thebibliography}{60}
\expandafter\ifx\csname natexlab\endcsname\relax\def\natexlab#1{#1}\fi
\expandafter\ifx\csname url\endcsname\relax
  \def\url#1{{\tt #1}}\fi
\expandafter\ifx\csname urlprefix\endcsname\relax\def\urlprefix{URL }\fi
\expandafter\ifx\csname urlstyle\endcsname\relax
  \expandafter\ifx\csname doi\endcsname\relax
  \def\doi#1{doi:\discretionary{}{}{}#1}\fi \else
  \expandafter\ifx\csname doi\endcsname\relax
  \def\doi{doi:\discretionary{}{}{}\begingroup \urlstyle{rm}\Url}\fi \fi

\bibitem[{Assi et~al.(2011)Assi, Brown, Djibo, Norman, Rajgopal, Welling, Chen,
  Bailey, Kone, Kenea et~al.}]{assi2011impact}
Assi, Tina-Marie, Shawn~T Brown, Ali Djibo, Bryan~A Norman, Jayant Rajgopal,
  Joel~S Welling, Sheng-I Chen, Rachel~R Bailey, Souleymane Kone, Hailu Kenea,
  et~al. 2011.
\newblock Impact of changing the measles vaccine vial size on {N}iger's vaccine
  supply chain: a computational model.
\newblock {\it BMC Public Health\/} {\bf 11}(1) 425.

\bibitem[{Assi et~al.(2013)Assi, Brown, Kone, Norman, Djibo, Connor, Wateska,
  Rajgopal, Slayton, and Lee}]{assi2013removing}
Assi, Tina-Marie, Shawn~T Brown, Souleymane Kone, Bryan~A Norman, Ali Djibo,
  Diana~L Connor, Angela~R Wateska, Jayant Rajgopal, Rachel~B Slayton, Bruce~Y
  Lee. 2013.
\newblock Removing the regional level from the {N}iger vaccine supply chain.
\newblock {\it Vaccine\/} {\bf 31}(26) 2828--2834.

\bibitem[{Azadi et~al.(2019)Azadi, Gangammanavar, and
  Eksioglu}]{azadi2019developing}
Azadi, Zahra, Harsha Gangammanavar, Sandra Eksioglu. 2019.
\newblock Developing childhood vaccine administration and inventory
  replenishment policies that minimize open vial wastage.
\newblock {\it Annals of Operations Research\/}  1--33.

\bibitem[{Azadi et~al.(2018)Azadi, Neyens, and Eksioglu}]{azadi2018iise}
Azadi, Zahra, David~M Neyens, Sandra~D Eksioglu. 2018.
\newblock Forecasting childhood routine immunization vaccine demand: A case
  study in {N}iger.
\newblock {\it IIE Annual Conference. Proceedings\/}. Institute of Industrial
  and Systems Engineers (IISE).

\bibitem[{Bakker et~al.(2012)Bakker, Riezebos, and Teunter}]{Bakker12}
Bakker, M., J.~Riezebos, R.~Teunter. 2012.
\newblock Review of inventory systems with deterioration since 2001.
\newblock {\it European Journal of Operational Research\/} {\bf 221} 275--284.

\bibitem[{Balcik et~al.(2015)Balcik, Haavisto, and
  Goentzel}]{balcik2015measuring}
Balcik, Burcu, Ira Haavisto, Jarrod Goentzel. 2015.
\newblock Measuring humanitarian supply chain performance in a multi-goal
  context.
\newblock {\it Journal of Humanitarian Logistics and Supply Chain Management\/}
  .

\bibitem[{Beamon and Balcik(2008)}]{beamon2008performance}
Beamon, Benita~M, Burcu Balcik. 2008.
\newblock Performance measurement in humanitarian relief chains.
\newblock {\it International Journal of Public Sector Management\/} {\bf 31}(1)
  4--25.

\bibitem[{Beraldi et~al.(2004)Beraldi, Bruni, and
  Conforti}]{beraldi2004designing}
Beraldi, Patrizia, Maria~Elena Bruni, Domenico Conforti. 2004.
\newblock Designing robust emergency medical service via stochastic
  programming.
\newblock {\it European Journal of Operational Research\/} {\bf 158}(1)
  183--193.

\bibitem[{Blackburn and Scudder(2009)}]{BlackburnScudder}
Blackburn, Joseph, Gary Scudder. 2009.
\newblock Supply chain strategies for perishable products: The case of fresh
  produce.
\newblock {\it Production and Operations Management\/} {\bf 18}(2) 129--137.

\bibitem[{Brown et~al.(2014)Brown, Schreiber, Cakouros, Wateska, Dicko, Connor,
  Jaillard, Mvundura, Norman, Levin et~al.}]{brown2014benefits}
Brown, Shawn~T, Benjamin Schreiber, Brigid~E Cakouros, Angela~R Wateska,
  Hamadou~M Dicko, Diana~L Connor, Philippe Jaillard, Mercy Mvundura, Bryan~A
  Norman, Carol Levin, et~al. 2014.
\newblock The benefits of redesigning benin's vaccine supply chain.
\newblock {\it Vaccine\/} {\bf 32}(32) 4097--4103.

\bibitem[{Chen et~al.(2014)Chen, Norman, Rajgopal, Assi, Lee, and
  Brown}]{chen2014planning}
Chen, Sheng-I, Bryan~A Norman, Jayant Rajgopal, Tina~M Assi, Bruce~Y Lee,
  Shawn~T Brown. 2014.
\newblock A planning model for the {WHO-EPI} vaccine distribution network in
  developing countries.
\newblock {\it IIE Transactions\/} {\bf 46}(8) 853--865.

\bibitem[{Chick et~al.(2008)Chick, Mamani, and Simchi-Levi}]{chick2008supply}
Chick, Stephen~E, Hamed Mamani, David Simchi-Levi. 2008.
\newblock Supply chain coordination and influenza vaccination.
\newblock {\it Operations Research\/} {\bf 56}(6) 1493--1506.

\bibitem[{Dai et~al.(2016)Dai, Cho, and Zhang}]{dai2016contracting}
Dai, Tinglong, Soo-Haeng Cho, Fuqiang Zhang. 2016.
\newblock Contracting for on-time delivery in the {US} influenza vaccine supply
  chain.
\newblock {\it Manufacturing \& Service Operations Management\/} {\bf 18}(3)
  332--346.

\bibitem[{Dave(1991)}]{Dave91}
Dave, U. 1991.
\newblock Survey of literature on continuously deteriorating inventory
  models—a rejoinder.
\newblock {\it Journal of the Operational Research Society\/} {\bf 42}(8) 725.

\bibitem[{Devore(2011)}]{devore2011probability}
Devore, Jay~L. 2011.
\newblock {\it Probability and Statistics for Engineering and the Sciences\/}.
\newblock Cengage learning.

\bibitem[{Dhamodharan and Proano(2012)}]{dhamodharan2012determining}
Dhamodharan, Aswin, Ruben~A Proano. 2012.
\newblock Determining the optimal vaccine vial size in developing countries: a
  monte carlo simulation approach.
\newblock {\it Healthcare Management Science\/} {\bf 15}(3) 188--196.

\bibitem[{Donkor and Duffey(2013)}]{donkor2013optimal}
Donkor, Emmanuel~A, Michael Duffey. 2013.
\newblock Optimal capital structure and financial risk of project finance
  investments: A simulation optimization model with chance constraints.
\newblock {\it The Engineering Economist\/} {\bf 58}(1) 19--34.

\bibitem[{Duijzer et~al.(2018{\natexlab{a}})Duijzer, van Jaarsveld, and
  Dekker}]{DuijzerSurvey}
Duijzer, L.E., W.~van Jaarsveld, R.~Dekker. 2018{\natexlab{a}}.
\newblock Literature review: The vaccine supply chain.
\newblock {\it European Journal of Operational Research\/} {\bf 268} 174--192.

\bibitem[{Duijzer et~al.(2018{\natexlab{b}})Duijzer, van Jaarsveld, Wallinga,
  and Dekker}]{duijzer2018dose}
Duijzer, Lotty~E, Willem~L van Jaarsveld, Jacco Wallinga, Rommert Dekker.
  2018{\natexlab{b}}.
\newblock Dose-optimal vaccine allocation over multiple populations.
\newblock {\it Production and Operations Management\/} {\bf 27}(1) 143--159.

\bibitem[{Galen(2019)}]{Palmetto}
Galen, C. 2019.
\newblock Palmetto cluster user guide.
\newblock \url{https://www.palmetto.clemson.edu/palmetto/index.html}.
\newblock Accessed: 2019-05-13.

\bibitem[{Gaulkler et~al.(2020)Gaulkler, Ketzenberg, and Zuidwijk}]{Gaulkler20}
Gaulkler, Gary, Michael Ketzenberg, Robert Zuidwijk. 2020.
\newblock The value of time and temperature history information for effective
  distribution of perishables.
\newblock {\it working paper\/} .

\bibitem[{Geletu et~al.(2013)Geletu, Kl{\"o}ppel, Zhang, and
  Li}]{geletu2013advances}
Geletu, Abebe, Michael Kl{\"o}ppel, Hui Zhang, Pu~Li. 2013.
\newblock Advances and applications of chance-constrained approaches to systems
  optimisation under uncertainty.
\newblock {\it International Journal of Systems Science\/} {\bf 44}(7)
  1209--1232.

\bibitem[{Govindan et~al.(2014)Govindan, Jafarian, Khodaverdi, and
  Devika}]{Govindan14}
Govindan, K., A.~Jafarian, R.~Khodaverdi, K.~Devika. 2014.
\newblock Two-echelon multiple-vehicle location–routing problem with time
  windows for optimization of sustainable supply chain network of perishable
  food.
\newblock {\it Int. J. Production Economics\/} {\bf 152} 9--28.

\bibitem[{Goyal and Giri(2001)}]{Goyal01}
Goyal, S., B.~Giri. 2001.
\newblock Recent trends in modeling of deteriorating inventory.
\newblock {\it European Journal of Operational Research\/} {\bf 134} 1--16.

\bibitem[{Haidari et~al.(2013)Haidari, Connor, Wateska, Brown, Mueller, Norman,
  Schmitz, Paul, Rajgopal, Welling et~al.}]{haidari2013augmenting}
Haidari, Leila~A, Diana~L Connor, Angela~R Wateska, Shawn~T Brown, Leslie~E
  Mueller, Bryan~A Norman, Michelle~M Schmitz, Proma Paul, Jayant Rajgopal,
  Joel~S Welling, et~al. 2013.
\newblock Augmenting transport versus increasing cold storage to improve
  vaccine supply chains.
\newblock {\it PLoS One\/} {\bf 8}(5) e64303.

\bibitem[{Haidari et~al.(2015)Haidari, Wahl, Brown, Privor-Dumm,
  Wallman-Stokes, Gorham, Connor, Wateska, Schreiber, Dicko
  et~al.}]{haidari2015one}
Haidari, Leila~A, Brian Wahl, Shawn~T Brown, Lois Privor-Dumm, Cecily
  Wallman-Stokes, Katie Gorham, Diana~L Connor, Angela~R Wateska, Benjamin
  Schreiber, Hamadou Dicko, et~al. 2015.
\newblock One size does not fit all: The impact of primary vaccine container
  size on vaccine distribution and delivery.
\newblock {\it Vaccine\/} {\bf 33}(28) 3242--3247.

\bibitem[{Henrion(2004)}]{henrion2004introduction}
Henrion, Ren{\'e}. 2004.
\newblock Introduction to chance-constrained programming.
\newblock {\it Tutorial paper for the Stochastic Programming Community\/} .

\bibitem[{Hovav and Tsadikovich(2015)}]{hovav2015network}
Hovav, Sharon, Dmitry Tsadikovich. 2015.
\newblock A network flow model for inventory management and distribution of
  influenza vaccines through a healthcare supply chain.
\newblock {\it Operations Research for Health Care\/} {\bf 5} 49--62.

\bibitem[{Humphreys(2011)}]{Humphreys11}
Humphreys, G. 2011.
\newblock Bulletin of the world health organization: Vaccination: rattling the
  supply chain.
\newblock
  \urlprefix\url{https://www.who.int/bulletin/volumes/89/5/11-030511/en/}.

\bibitem[{ICF(2017)}]{DHS}
ICF, Funded by~USAID. 2017.
\newblock Demographic health survey
  \urlprefix\url{https://dhsprogram.com/data/data-collection.cfm}.

\bibitem[{Jiang and Guan(2016)}]{jiang2016data}
Jiang, Ruiwei, Yongpei Guan. 2016.
\newblock Data-driven chance constrained stochastic program.
\newblock {\it Mathematical Programming\/} {\bf 158}(1-2) 291--327.

\bibitem[{Karaesmen et~al.(2011)Karaesmen, Scheller-Wolf, and
  Deniz}]{karaesmen2011managing}
Karaesmen, Itir~Z, Alan Scheller-Wolf, Borga Deniz. 2011.
\newblock Managing perishable and aging inventories: review and future research
  directions.
\newblock {\it Planning production and inventories in the extended
  enterprise\/}. Springer, 393--436.

\bibitem[{Ketzenberg et~al.(2015)Ketzenberg, Bloemhof, and Gaulkler}]{Ketz15}
Ketzenberg, Michael, Jacqueline Bloemhof, Gary Gaulkler. 2015.
\newblock Managing perishables with time and temperature history.
\newblock {\it Production and Operations Management\/} {\bf 24}(1) 54--70.

\bibitem[{Kim et~al.(2015)Kim, Pasupathy, and Henderson}]{kim2015guide}
Kim, Sujin, Raghu Pasupathy, Shane~G Henderson. 2015.
\newblock A guide to sample average approximation.
\newblock {\it Handbook of Simulation Optimization\/}. Springer, 207--243.

\bibitem[{Kraiselburd and Yadav(2012)}]{KraiselYadav12}
Kraiselburd, Santiago, Prashant Yadav. 2012.
\newblock Supply chains and global health: An imperative for bringing
  operations management scholarship into action.
\newblock {\it Production and Operations Management\/} {\bf 22}(2) 377--381.

\bibitem[{K{\"u}{\c{c}}{\"u}kyavuz(2012)}]{kuccukyavuz2012mixing}
K{\"u}{\c{c}}{\"u}kyavuz, Simge. 2012.
\newblock On mixing sets arising in chance-constrained programming.
\newblock {\it Mathematical Programming\/} {\bf 132}(1-2) 31--56.

\bibitem[{Kurata(2014)}]{Kurata14}
Kurata, Hisashi. 2014.
\newblock How does inventory pooling work when product availability influences
  customers’ purchasing decisions?
\newblock {\it International Journal of Production Research\/} {\bf 52}(22)
  6739--6759.

\bibitem[{Lee et~al.(2012)Lee, Cakouros, Assi, Connor, Welling, Kone, Djibo,
  Wateska, Pierre, and Brown}]{lee2012impact}
Lee, Bruce~Y, Brigid~E Cakouros, Tina-Marie Assi, Diana~L Connor, Joel Welling,
  Souleymane Kone, Ali Djibo, Angela~R Wateska, Lionel Pierre, Shawn~T Brown.
  2012.
\newblock The impact of making vaccines thermostable in {N}iger's vaccine
  supply chain.
\newblock {\it Vaccine\/} {\bf 30}(38) 5637--5643.

\bibitem[{Lee et~al.(2015)Lee, Connor, Wateska, Norman, Rajgopal, Cakouros,
  Chen, Claypool, Haidari, Karir et~al.}]{lee2015landscaping}
Lee, Bruce~Y, Diana~L Connor, Angela~R Wateska, Bryan~A Norman, Jayant
  Rajgopal, Brigid~E Cakouros, Sheng-I Chen, Erin~G Claypool, Leila~A Haidari,
  Veena Karir, et~al. 2015.
\newblock Landscaping the structures of {GAVI} country vaccine supply chains
  and testing the effects of radical redesign.
\newblock {\it Vaccine\/} {\bf 33}(36) 4451--4458.

\bibitem[{Lee et~al.(2016)Lee, Haidari, Prosser, Connor, Bechtel, Dipuve,
  Kassim, Khanlawia, and Brown}]{lee2016re}
Lee, Bruce~Y, Leila~A Haidari, Wendy Prosser, Diana~L Connor, Ruth Bechtel,
  Amelia Dipuve, Hidayat Kassim, Balbina Khanlawia, Shawn~T Brown. 2016.
\newblock Re-designing the {M}ozambique vaccine supply chain to improve access
  to vaccines.
\newblock {\it Vaccine\/} {\bf 34}(41) 4998--5004.

\bibitem[{Lee et~al.(2017)Lee, Wedlock, Haidari, Elder, Potet, Manring, Connor,
  Spiker, Bonner, Rangarajan et~al.}]{lee2017economic}
Lee, Bruce~Y, Patrick~T Wedlock, Leila~A Haidari, Kate Elder, Julien Potet,
  Rachel Manring, Diana~L Connor, Marie~L Spiker, Kimberly Bonner, Arjun
  Rangarajan, et~al. 2017.
\newblock Economic impact of thermostable vaccines.
\newblock {\it Vaccine\/} {\bf 35}(23) 3135--3142.

\bibitem[{Li et~al.(2010)Li, Lan, and Mawhinney}]{LiLanMaw10}
Li, R., H.~Lan, J.~Mawhinney. 2010.
\newblock A review on deteriorating inventory study.
\newblock {\it Journal of Service Science and Management\/} {\bf 3}(1)
  117--129.

\bibitem[{Lubin et~al.(2016)Lubin, Dvorkin, and Backhaus}]{lubin2016robust}
Lubin, Miles, Yury Dvorkin, Scott Backhaus. 2016.
\newblock A robust approach to chance constrained optimal power flow with
  renewable generation.
\newblock {\it IEEE Transactions on Power Systems\/} {\bf 31}(5) 3840--3849.

\bibitem[{Luedtke and Ahmed(2008)}]{luedtke2008sample}
Luedtke, James, Shabbir Ahmed. 2008.
\newblock A sample approximation approach for optimization with probabilistic
  constraints.
\newblock {\it SIAM Journal on Optimization\/} {\bf 19}(2) 674--699.

\bibitem[{McCoy and Lee(2014)}]{mccoy2014using}
McCoy, Jessica~H, Hau~L Lee. 2014.
\newblock Using fairness models to improve equity in health delivery fleet
  management.
\newblock {\it Production and Operations Management\/} {\bf 23}(6) 965--977.

\bibitem[{Mueller et~al.(2016)Mueller, Haidari, Wateska, Phillips, Schmitz,
  Connor, Norman, Brown, Welling, and Lee}]{mueller2016impact}
Mueller, Leslie~E, Leila~A Haidari, Angela~R Wateska, Roslyn~J Phillips,
  Michelle~M Schmitz, Diana~L Connor, Bryan~A Norman, Shawn~T Brown, Joel~S
  Welling, Bruce~Y Lee. 2016.
\newblock The impact of implementing a demand forecasting system into a
  low-income country's supply chain.
\newblock {\it Vaccine\/} {\bf 34}(32) 3663--3669.

\bibitem[{Nagurney et~al.(2010)Nagurney, Yu, Masoumi, and
  Nagurney}]{NagurneyYu10}
Nagurney, Anna, Min Yu, Amir~H. Masoumi, Ladimer~S Nagurney. 2010.
\newblock {\it Networks Against Time: Supply Chain Analytics for Perishable
  Products\/}.
\newblock Springer: New York.

\bibitem[{Nahmias(1982)}]{Nahmias82}
Nahmias, S. 1982.
\newblock Perishable inventory theory: A review.
\newblock {\it Operations Research\/} {\bf 30}(4) 680--707.

\bibitem[{Nemirovski(2012)}]{nemirovski2012safe}
Nemirovski, Arkadi. 2012.
\newblock On safe tractable approximations of chance constraints.
\newblock {\it European Journal of Operational Research\/} {\bf 219}(3)
  707--718.

\bibitem[{Pagnoncelli et~al.(2009)Pagnoncelli, Ahmed, and
  Shapiro}]{pagnoncelli2009sample}
Pagnoncelli, BK, Shapiro Ahmed, Alexander Shapiro. 2009.
\newblock Sample average approximation method for chance constrained
  programming: theory and applications.
\newblock {\it Journal of optimization theory and applications\/} {\bf 142}(2)
  399--416.

\bibitem[{Raafat(1991)}]{Raafat}
Raafat, F. 1991.
\newblock Survey of literature on continuously deteriorating inventories.
\newblock {\it Journal of the Operational Research Society\/} {\bf 42}(1)
  27--37.

\bibitem[{Riewpaiboon et~al.(2015)Riewpaiboon, Sooksriwong, Chaiyakunapruk,
  Tharmaphornpilas, Techathawat, Rookkapan, Rasdjarmrearnsook, and
  Suraratdecha}]{riewpaiboon2015optimizing}
Riewpaiboon, Arthorn, Chaoncin Sooksriwong, Nathorn Chaiyakunapruk, Piyanit
  Tharmaphornpilas, Sirirat Techathawat, Korpong Rookkapan, Aim-orn
  Rasdjarmrearnsook, Chutima Suraratdecha. 2015.
\newblock Optimizing national immunization program supply chain management in
  {T}hailand: an economic analysis.
\newblock {\it Public Health\/} {\bf 129}(7) 899--906.

\bibitem[{Van~Ackooij et~al.(2011)Van~Ackooij, Zorgati, Henrion, and
  M{\"o}ller}]{van2011chance}
Van~Ackooij, Wim, Riadh Zorgati, Ren{\'e} Henrion, Andris M{\"o}ller. 2011.
\newblock Chance constrained programming and its applications to energy
  management.
\newblock {\it Stochastic Optimization—Seeing the Optimal for the
  Uncertain\/}  291--320.

\bibitem[{Wedlock et~al.(2018)Wedlock, Mitgang, Siegmund, DePasse, Bakal,
  Leonard, Welling, Brown, and Lee}]{wedlock2018dual}
Wedlock, Patrick~T, Elizabeth~A Mitgang, Sheryl~S Siegmund, Jay DePasse,
  Jennifer Bakal, Jim Leonard, Joel Welling, Shawn~T Brown, Bruce~Y Lee. 2018.
\newblock Dual-chamber injection device for measles-rubella vaccine: The
  potential impact of introducing varying sizes of the devices in 3 countries.
\newblock {\it Vaccine\/} {\bf 36}(39) 5879--5885.

\bibitem[{Westerink-Duijzer et~al.(2020)Westerink-Duijzer, Schlicher, and
  Musegaas}]{f1fbee18d195400980a3bc34a3f10f71}
Westerink-Duijzer, Evelot, {Loe P.J.} Schlicher, Marieke Musegaas. 2020.
\newblock Core allocations for cooperation problems in vaccination.
\newblock {\it Production and Operations Management\/} .

\bibitem[{WHO(1974)}]{WHOBenefit}
WHO. 1974.
\newblock The expanded programme on immunization.
\newblock
  \url{http://www.who.int/immunization/programmes_systems/supply_chain/benefits_of_immunization/en/}.
\newblock Online; accessed May 2018.

\bibitem[{{WHO}(2014)}]{WHOImmunization}
{WHO}. 2014.
\newblock {Immunization coverage reaches 84\%, still short of 90\% goal }.
\newblock
  \url{http://www.who.int/immunization/newsroom/press/immunization_coverage_july2014/en/}.
\newblock Online; accessed May 2018.

\bibitem[{WHO(2019{\natexlab{a}})}]{WHO20mill}
WHO. 2019{\natexlab{a}}.
\newblock 20 million children miss out on lifesaving measles, diphtheria and
  tetanus vaccines in 2018.
\newblock \urlprefix\url{https://www.who.int/news-room/detail/}.

\bibitem[{WHO(2019{\natexlab{b}})}]{WHO2018Facts}
WHO. 2019{\natexlab{b}}.
\newblock Children: reducing mortality.
\newblock
  \url{https://www.who.int/news-room/fact-sheets/detail/children-reducing-mortality}.

\bibitem[{Zhou et~al.(2011)Zhou, Leung, and Pierskalla}]{Zhou11}
Zhou, Deming, Lawrence~C. Leung, William~P. Pierskalla. 2011.
\newblock Inventory management of platelets in hospitals: Optimal inventory
  policy for perishable products with regular and optional expedited
  replenishments.
\newblock {\it Manufacturing \& Service Operations Management\/} {\bf 13}(4)
  420--438.

\end{thebibliography}

\begin{appendices}
\section{Chance-constrained programming model} \label{appA}
The problem parameters and decision variables are presented as follows:
\begin{figure}{h}
    \centering
    \includegraphics[width=\textwidth]{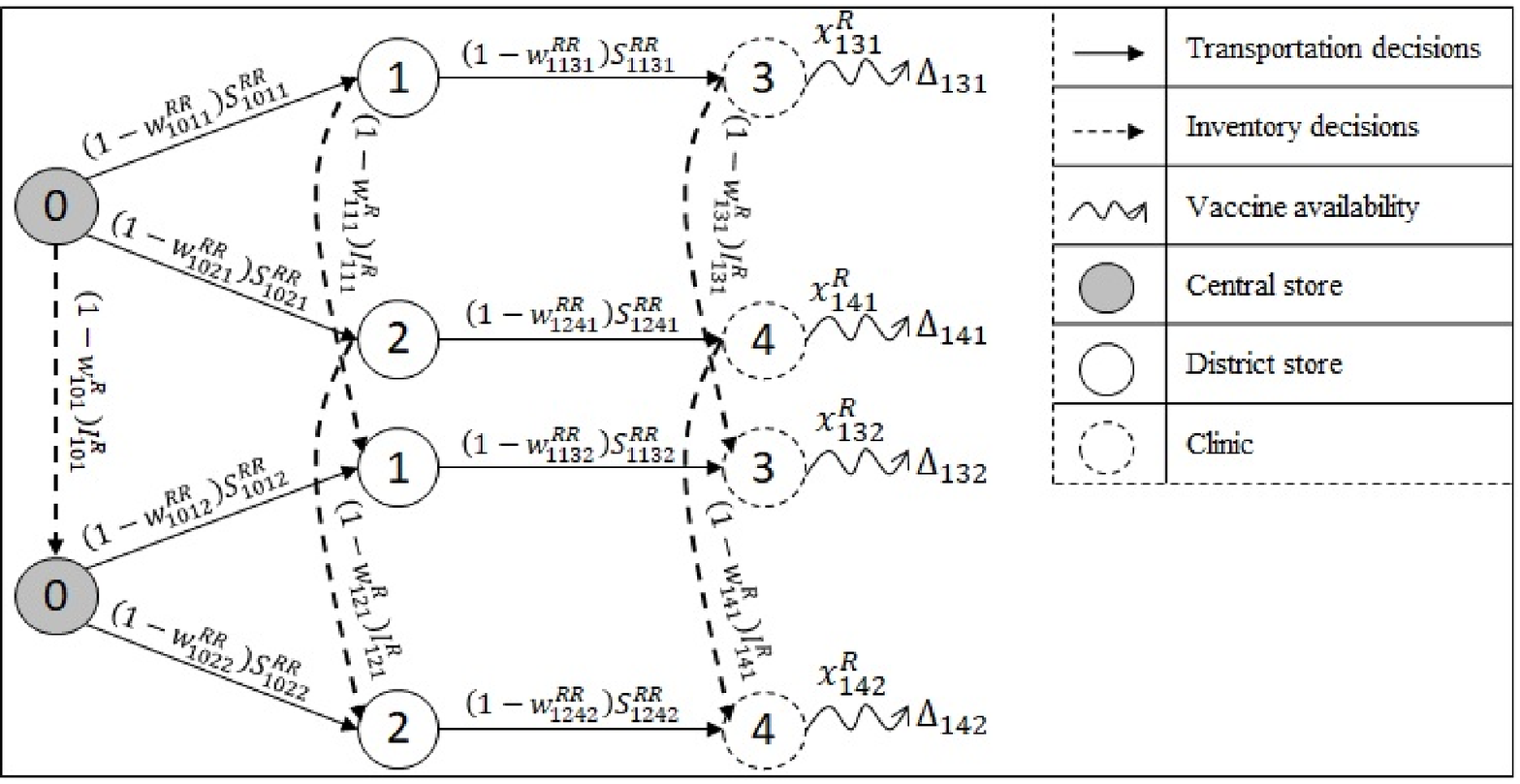}
    \centering
    \caption{A vaccine supply chain distribution network with three tiers: 1 central store, 2 district stores, and 1 clinic.} \label{fig-Network}
\end{figure}
\figref{fig-Network} provides a  schematic representation of our modeling a three-tier vaccine supply chain with one central store, two district stores, and one clinic that uses only refrigerators for storing vaccines. This network considers one vaccine type and two time periods. The nodes in successive time periods are connected to represent the flow of inventory. Inventory arcs are represented by the dotted lines. The solid arcs represent transportation decisions.

\textbf{Set of Indices}:  \\
$\mathcal{I}$: Set of vaccine types. \\
$\mathcal{I}^R$: Set of vaccines that can only be stored in a refrigerator. $\mathcal{I}^R \subset \mathcal{I}$. \\
$\mathcal{J}$: Set of the nodes in the network. \\
$\mathcal{T}$: Set of periods in the planning horizon.

\textbf{Parameters}:\\
$\mu_{ijt}$: Mean demand for doses of vaccine type $i$ at location $j$ in time period $t$. \\
$\mu'_{ij}$: Mean daily demand for doses of vaccine type $i$ at location $j$. \\
${\omega}_{ijt}$: Uncertain parameter representing the demand for doses of vaccine type $i$ at location $j$ in time period $t$. \\
$C^R_j$: Effective refrigerator capacity at location $j$. \\
$C^F_j$: Effective freezer capacity at location $j$. \\
$C^V_{kj}$: Effective transport capacity from location $k$ to location $j$. \\
$q_i$: Effective packed volume of one dose of vaccine $i$. \\
$r_i$: Diluent volume for vaccine $i$. \\
$a_{i}$: Number of doses administered of vaccine $i$ within the vaccine regimen. \\
$\beta$: Risk factor.\\
$w^R_{ijt}$: Fraction of vaccine $i$ inventory in refrigerators lost at location $j$ in period $t$. \\
$w^F_{ijt}$: Fraction of vaccine $i$ inventory in freezers lost at location $j$ in period $t$. \\
$w^{RR}_{ikjt}$: Fraction of vaccine $i$ going from a refrigerator at location $k$ to a refrigerator at location $j$ in time period $t$ that is lost. \\
$w^{RF}_{ikjt}$: Fraction of vaccine $i$ going from a refrigerator at location $k$ to a freezer at location $j$ in time period $t$ that is lost. \\
$w^{FR}_{ikjt}$: Fraction of vaccine $i$ going from a freezer at location $k$ to a refrigerator at location $j$ in time period $t$ that is lost. \\
$w^{FF}_{ikjt}$: Fraction of vaccine $i$ going from a freezer at location $k$ to a freezer at location $j$ in time period $t$ that is lost. \\
$w^{O}_{ijt}$: Fraction of OVW for vaccine $i$ at location $j$ in time period $t$. \\
$\varepsilon$: Positive constant used to normalize the second term in the objective.

\textbf{Decision variables}: \\
$x^{R}_{ijt}$: Units of vaccine $i$ from a refrigerator that satisfy demand at location $j$ in period $t$. \\
$x^F_{ijt}$: Units of vaccine $i$ from a freezer that satisfy demand at location $j$ in period $t$. \\
$n_j$: Number of FIC at location $j$. \\
$I^R_{ijt}$: Inventory of vaccine $i$ in a refrigerator at location $j$ at the end of time period $t$. \\
$I^F_{ijt}$: Inventory of vaccine $i$ in a freezer at location $j$ at the end of time period $t$. \\
$S^{RR}_{ikjt}$: Units of vaccine $i$ shipped from a refrigerator at location $k$ to a refrigerator at location $j$ in time period $t$. \\
$S^{RF}_{ikjt}$: Units of vaccine $i$ shipped from a refrigerator at location $k$ to a freezer at location $j$ in time period $t$. \\
$S^{FR}_{ikjt}$: Units of vaccine $i$ shipped from a freezer at location $k$ to a refrigerator at location $j$ in time period $t$. \\
$S^{FF}_{ikjt}$: Units of vaccine $i$ shipped from a freezer at location $k$ to a freezer at location $j$ in time period $t$.

A planning model for the vaccine distribution network is formulated as the following chance-constrained program:

\begin{subequations}\label{Model_ASIS}
\begin{align}{4}
\max & \sum_{j \in \mathcal{J}} n_j +\varepsilon \sum_{i \in \mathcal{I}} \sum_{j \in \mathcal{J}}\sum_{t \in \mathcal{T}} (x^R_{ijt}+x^F_{ijt}), \label{OBJ-Original} \\
& s.t.\notag\\
&I^R_{ijt} = (1-w^R_{ijt-1})I^R_{ijt-1} +\sum_{k \in \mathcal{J},k\not=j}(1-w^{FR}_{ikjt-1})S^{FR}_{ikjt-1}+\sum_{k \in \mathcal{J},k\not=j}(1-w^{RR}_{ikjt-1})S^{RR}_{ikjt-1} \notag \\
&-\sum_{k \in \mathcal{J},k\not=j}S^{RF}_{ijkt} - \sum_{k \in \mathcal{J},k\not=j}S^{RR}_{ijkt} -\frac{x^{R}_{ijt}}{ (1-w^{O}_{ijt})}\quad \quad \forall i \in \mathcal{I}, j \in \mathcal{J}, t \in \mathcal{T}\setminus \{0\}, \label{Con-RefrigInv}\\
&I^F_{ijt} = (1-w^F_{ijt-1})I^F_{ijt-1} +\sum_{k \in \mathcal{J},k\not=j}(1-w^{RF}_{ikjt-1})S^{RF}_{ikjt-1}+\sum_{k \in \mathcal{J},k\not=j}(1-w^{FF}_{ikjt-1})S^{FF}_{ikjt-1} \notag \\
&-\sum_{k \in \mathcal{J},k\not=j}S^{FR}_{ijkt} - \sum_{k \in \mathcal{J},k\not=j}S^{FF}_{ijkt} - \frac{x^{F}_{ijt}}{(1-w^{O}_{ijt})}\quad \quad \forall i \in \mathcal{I}, j \in \mathcal{J}, t \in \mathcal{T}\setminus \{0\}, \label{Con-FreezInv}\\
&\sum_{i \in \mathcal{I}} q_i \bigg( I^R_{ijt}+ \sum_{k \in \mathcal{J},k\not=j}(1-w^{FR}_{ikjt})S^{FR}_{ikjt}+\sum_{k \in \mathcal{J},k\not=j}(1-w^{RR}_{ikjt})S^{RR}_{ikjt} \bigg) \notag \\
&\leq C^R_j - \sum_{i \in \mathcal{I}} r_i\mu'_{ij} \quad \quad \forall j \in \mathcal{J}, t \in \mathcal{T}, \label{Con-RefrigNodeCap}\\
&\sum_{i \in \mathcal{I}} q_i \bigg( I^F_{ij}+ \sum_{k \in \mathcal{J},k\not=j}(1-w^{RF}_{ikjt})S^{RF}_{ikjt}+\sum_{k \in \mathcal{J},k\not=j}(1-w^{FF}_{ikjt})S^{FF}_{ikjt} \bigg) \leq C^F_j \quad \quad \forall j \in \mathcal{J}, t \in \mathcal{T}, \label{Con-FreezNodeCap}\\
&I^R_{ij0} = 0 \quad \quad \forall i \in \mathcal{I}, j \in \mathcal{J}, \label{Con-RefrigInitInv}\\
&I^F_{ij0} = 0 \quad \quad \forall i \in \mathcal{I}, j \in \mathcal{J}, \label{Con-FreezInitInv}\\
&I^F_{ij|T|} = 0 \quad \quad \forall i \in \mathcal{I}^{R}, j \in \mathcal{J}, \label{Con-FreezNoStore}\\
&\sum_{i \in \mathcal{I}} q_i (S^{RR}_{ikjt}+S^{RF}_{ikjt}+S^{FR}_{ikjt}+S^{FF}_{ikjt}) \leq C^V_{kj} \quad \quad \forall j,k \in \mathcal{J};j\neq k, t \in \mathcal{T}, \label{Con-ArcCap}\\
&n_j  \leq \sum_{t \in \mathcal{T}}(x^R_{ijt}+x^F_{ijt})/a_i \quad \quad \forall i \in \mathcal{I}, j \in \mathcal{J}, \label{Con-FIC}\\
&Pr \bigg(x^R_{ijt}+x^F_{ijt} \geq \Delta_{ijt} \bigg)\geq \beta \quad \quad \forall i \in \mathcal{I}, j \in \mathcal{J},t \in \mathcal{T}, \label{Con-DemandLeq-extendedmodel} \\
& x^{R/F}_{ijt}, n_j, I^{R/F}_{ijt}, S^{RR/RF/FR/FF}_{ikjt}  \geq 0 \quad\quad \forall i \in \mathcal{I}, j,k \in \mathcal{J};j \not = k, t \in \mathcal{T}.  \label{Con_nonnegativity}
\end{align}
\end{subequations}

In this model, the first term in the objective function maximizes the percentage of FIC over all clinics. This results in maximizing the number of children who receive the required doses specified by the vaccine regimen.
The second term presents the total number of doses delivered to clinics, which impacts SR. Since these two terms have different units, we normalize the second term by multiplying it with a coefficient $\varepsilon \in [0,1]$. By varying the value of this coefficient one can give more or less relative weight to maximizing the percentage of FIC.

Equations \eqref{Con-RefrigInv} and \eqref{Con-FreezInv} are the inventory balance constraints for refrigerators and freezers at each facility in the supply chain. It is assumed that the lead time for shipping vaccines between successive tiers of the supply chain, is one week. The model accounts for specific shipment schedules between locations $j$ and $k$ by fixing variables $S^{RR}_{ikjt}$, $S^{RF}_{ikjt}$, $S^{FR}_{ikjt}$, and $S^{FF}_{ikjt}$ to $0$ during time periods when there are no shipments. Doing this allows us to model realistic situations, such as a schedule that delivers vaccines from the central to regional stores every three months.

Constraints \eqref{Con-RefrigNodeCap} and \eqref{Con-FreezNodeCap} limit the storage capacity of refrigerators and freezers, respectively. We adjust the refrigerator capacity in \eqref{Con-RefrigNodeCap} to account for the fact that diluents are refrigerated before vaccination. Constraints \eqref{Con-RefrigInitInv} and \eqref{Con-FreezInitInv} initialize the inventory of vaccines in refrigerators and freezers, respectively. Constraints \eqref{Con-FreezNoStore} guarantee that vaccines that are not stable at freezer temperature must only be stored in refrigerators. Constraint \eqref{Con-ArcCap} limits the transportation capacity between two locations. Note that the model assumes both freezable and non-freezable vaccines are shipped via the same vehicle type. Constraint \eqref{Con-FIC} determines FIC for each location. The right hand side of this inequality represents the average number of individuals to whom a full set of doses of vaccine $i$ is administered at clinic $j$  over the planning horizon, which is found by dividing the total number of vaccine $i$ doses administered by the number of required doses defined in the vaccine regimen at clinic $j$. Therefore, FIC is bounded by the smallest value among all vaccine types.

Finally, \eqref{Con-DemandLeq-extendedmodel} is a chance constraint that indicates the total number of vaccine type $i$ delivered to clinic $j$ in period $t$ should be greater than the total demand for vaccines at least $\beta*100\%$ of the time. This constraint captures the random nature of patient arrivals. This constraint also indicates that some times, the children present at a clinic are not  vaccinated if there is no inventory or if the session size is small. 
\newpage
\section{BS-SAA algorithm} \label{appB}
\begin{algorithm}[h]
  \caption{BS-SAA Algorithm} \label{SAA-Pseudocode}
  \textbf{Notation}: Let $\pi^{L}_{ijt}$ and $\pi^{U}_{ijt}$ denote the lower and upper bounds of the penalty term $\pi_{ijt}$. Also, let $\epsilon$ and $\vartheta$ be small positive constants ($\delta = \sigma = 0.01$).
  \begin{algorithmic}[1]
    \While{$|\pi_{ijt}-\frac{\pi_{ijt}^L+\pi_{ijt}^U}{2}| > \vartheta ~~ \forall i \in \mathcal{I}, j \in J, t \in \mathcal{T}$ }
        \State     \strut$\pi_{ijt} \gets \frac{\pi^L_{ijt}+\pi^U_{ijt}}{2}$
        \State Solve ($\bar{P}$) and let $\hat{v}^s_{ijt}$, be the incumbent solution of $v^s_{ijt}$, $\forall i \in \mathcal{I}, j \in J, t \in \mathcal{T}$.
     \For{$i \in \mathcal{I}, j \in J, t \in \mathcal{T}$}
            \State    \strut$M_{ijt} \gets 0$
        \For{$s=1,\dots,S$}
            \If {$\hat{v}^s_{ijt} < 0$}
                \State    \strut$M_{ijt} \gets M_{ijt}+1$
            \EndIf
        \EndFor
        \If {$M_{ijt} \leq (\ceil {1-\hat{\beta}}S)+ \epsilon$}
            \State     \strut$\pi^L_{ijt} \gets \frac{\pi^L_{ijt}+\pi^U_{ijt}}{2}$
        \ElsIf{ $M_{ijt} \geq (\ceil {1-\hat{\beta}}S) - \epsilon$}
            \State     \strut$\pi^U_{ijt} \gets \frac{\pi^L_{ijt}+\pi^U_{ijt}}{2}$
        \EndIf
    \EndFor
    \EndWhile
\State  Return $\pi_{ijt}, ~ \forall i \in \mathcal{I}, j \in J, t \in \mathcal{T}$, and the solution to ($\bar{P}$)
  \end{algorithmic}
\end{algorithm}

\newpage
\section{Extended chance-constrained programming model} \label{appC}
To extend model in \eqref{Model_ASIS}, a new set $\mathcal{N}^i$ is introduced to denote different sizes of the vaccine type $i$. Note that vial size refers to the number of doses per vial. Moreover, a superscript $\iota$ is introduced to denote the size of the vaccine.

The new problem parameters and decision variables are presented as follows:

\textbf{Set of Indices}:  \\
${I}$: Set of vaccine types. \\
$\mathcal{N}^i$: Set of available vial sizes for vaccines type $i$. \\
$\mathcal{I}^R$: Set of vaccines that can only be stored in a refrigerator. $\mathcal{I}^R \subset (\mathcal{I} \times  \mathcal{N}^i)$. \\
$\mathcal{J}$: Set of the nodes in the network. \\
$\mathcal{T}$: Set of periods in the planning horizon.

\textbf{Parameters}:\\
$\mu_{ijt}$: Mean demand for doses of vaccine type $i$ at location $j$ in time period $t$. \\
$\mu'_{ij}$: Mean daily demand for doses of vaccine type $i$ at location $j$. \\
${\omega}_{ijt}$: Uncertain parameter which represents the demand for doses of vaccine type $i$ at location $j$ in time period $t$. \\
$C^R_j$: Effective refrigerator capacity at location $j$. \\
$C^F_j$: Effective freezer capacity at location $j$. \\
$C^V_{kj}$: Effective transport capacity from location $k$ to location $j$. \\
$q_{i}^{\iota}$: Effective packed volume of one dose of vaccine type $i$ of vial size $\iota$. \\
$r_i^\iota$: Diluent volume for vaccine type $i$ of vial size $\iota$. \\
$a_{i}$: Number of doses administered of vaccine type $i$ within the vaccine regimen. \\
$b_{i}^{\iota}$: Number of doses per vial for vaccine type $i$ of size $\iota$. \\
$\beta$: Risk factor.\\
$w^{R\iota}_{ijt}$: Fraction of vaccine type $i$ of size $\iota$ inventory in refrigerators lost at location $j$ in period $t$. \\
$w^{F\iota}_{ijt}$: Fraction of vaccine type $i$ of size $\iota$ inventory in freezers lost at location $j$ in period $t$. \\
$w^{RR\iota}_{ikjt}$: Fraction of vaccine type $i$ of size $\iota$ going from a refrigerator at location $k$ to a refrigerator at location $j$ in time period $t$ that is lost. \\
$w^{RF\iota}_{ikjt}$: Fraction of vaccine type $i$ of size $\iota$ going from a refrigerator at location $k$ to a freezer at location $j$ in time period $t$ that is lost. \\
$w^{FR\iota}_{ikjt}$: Fraction of vaccine type $i$ of size $\iota$ going from a freezer at location $k$ to a refrigerator at location $j$ in time period $t$ that is lost. \\
$w^{FF\iota}_{ikjt}$: Fraction of vaccine type $i$ of size $\iota$ going from a freezer at location $k$ to a freezer at location $j$ in time period $t$ that is lost. \\
$w^{O\iota}_{ijt}$: Fraction of OVW for vaccine type $i$ of size $\iota$ at location $j$ in time period $t$. \\
$\varepsilon$: Positive constant used to normalize the second term in the objective.

\textbf{Decision variables}: \\
$x^{R\iota}_{ijt}$: Units of vaccine type $i$ of size $\iota$ used from a refrigerator to satisfy demand at location $j$ in period $t$. \\
$x^{F\iota}_{ijt}$: Units of vaccine type $i$ of size $\iota$ used from a freezer to satisfy demand at location $j$ in period $t$. \\
$n_j$: Number of FIC at location $j$. \\
$I^{R\iota}_{ijt}$: Inventory of vaccine type $i$ of size $\iota$ in a refrigerator at location $j$ at end of time period $t$. \\
$I^{F\iota}_{ijt}$: Inventory of vaccine type $i$ of size $\iota$ in a freezer at location $j$ at end of time period $t$. \\
$S^{RR\iota}_{ikjt}$: Units of vaccine type $i$ of size $\iota$ shipped from a refrigerator at location $k$ to a refrigerator at location $j$ in time period $t$. \\
$S^{RF\iota}_{ikjt}$: Units of vaccine type $i$ of size $\iota$ shipped from a refrigerator at location $k$ to a freezer at location $j$ in time period $t$. \\
$S^{FR\iota}_{ikjt}$: Units of vaccine type $i$ of size $\iota$ shipped from a freezer at location $k$ to a refrigerator at location $j$ in time period $t$. \\
$S^{FF\iota}_{ikjt}$: Units of vaccine type $i$ of size $\iota$ shipped from a freezer at location $k$ to a freezer at location $j$ in time period $t$.

Here is the extended model formulated for the chance-constrained program in Appendix \ref{appA}:

\begin{subequations}\label{Model_Extended}
\begin{align}
\max & \sum_{j \in \mathcal{J}} n_j +\varepsilon \sum_{i \in \mathcal{I}} \sum_{j \in \mathcal{J}}\sum_{t \in \mathcal{T}} \big ( \sum_{\iota \in \mathcal{N}^i} (x^{R\iota}_{ijt}+x^{F\iota}_{ijt}) \big ), \label{OBJ-_Extended} \\
& s.t.\notag\\
&I^{R\iota}_{ijt} = (1-w^{R\iota}_{ijt-1})I^{R\iota}_{ijt-1} +\sum_{k \in \mathcal{J},k\not=j}(1-w^{FR\iota}_{ikjt-1})S^{FR\iota}_{ikjt-1}+\sum_{k \in \mathcal{J},k\not=j}(1-w^{RR\iota}_{ikjt-1})S^{RR\iota}_{ikjt-1} \notag \\
&-\sum_{k \in \mathcal{J},k\not=j}S^{RF\iota}_{ijkt} - \sum_{k \in \mathcal{J},k\not=j}S^{RR\iota}_{ijkt} -\frac{x^{R\iota}_{ijt}}{ (1-w^{O\iota}_{ijt})}\quad \quad \forall i \in \mathcal{I}, \iota \in \mathcal{N}^i, j \in \mathcal{J}, t \in \mathcal{T}\setminus \{0\}, \label{ConExtended-RefrigInv}\\
&I^{F\iota}_{ijt} = (1-w^{F\iota}_{ijt-1})I^{F\iota}_{ijt-1} +\sum_{k \in \mathcal{J},k\not=j}(1-w^{RF\iota}_{ikjt-1})S^{RF\iota}_{ikjt-1}+\sum_{k \in \mathcal{J},k\not=j}(1-w^{FF\iota}_{ikjt-1})S^{FF\iota}_{ikjt-1} \notag \\
&-\sum_{k \in \mathcal{J},k\not=j}S^{FR\iota}_{ijkt} - \sum_{k \in \mathcal{J},k\not=j}S^{FF\iota}_{ijkt} - \frac{x^{F\iota}_{ijt}}{(1-w^{O\iota}_{ijt})}\quad \quad \forall i \in \mathcal{N}, \iota \in \mathcal{N}^i, j \in \mathcal{J}, t \in \mathcal{T}\setminus \{0\}, \label{ConExtended-FreezInv}\\
&\sum_{i \in \mathcal{I}}\sum_{\iota \in \mathcal{N}^i} q_i^\iota b_i^\iota \bigg( I^{R\iota}_{ijt}+ \sum_{k \in \mathcal{J},k\not=j}(1-w^{FR\iota}_{ikjt})S^{FR\iota}_{ikjt}+\sum_{k \in \mathcal{J},k\not=j}(1-w^{RR\iota}_{ikjt})S^{RR\iota}_{ikjt} \bigg) \notag \\
& \leq C^R_j - \sum_{i \in \mathcal{I}} r_i\mu'_{ij} \quad \quad \forall j \in \mathcal{J}, t \in \mathcal{T}, \label{ConExtended-RefrigNodeCap}\\
&\sum_{i \in \mathcal{I}}\sum_{\iota \in \mathcal{N}^i} q_i^\iota b_i^\iota \bigg( I^{F\iota}_{ij}+ \sum_{k \in \mathcal{J},k\not=j}(1-w^{RF\iota}_{ikjt})S^{RF\iota}_{ikjt}+\sum_{k \in \mathcal{J},k\not=j}(1-w^{FF\iota}_{ikjt})S^{FF\iota}_{ikjt} \bigg) \leq C^F_j \quad \quad \forall j \in \mathcal{J}, t \in \mathcal{T}, \label{ConExtended-FreezNodeCap}\\
&I^{R\iota}_{ij0} = 0 \quad \quad \forall i \in\mathcal{I}, \iota \in \mathcal{N}^i, j \in \mathcal{J}, \label{ConExtended-RefrigInitInv}\\
&I^{F\iota}_{ij0} = 0 \quad \quad \forall i \in \mathcal{I}, \iota \in \mathcal{N}^i, j \in \mathcal{J}, \label{ConExtended-FreezInitInv}\\
&I^{F\iota}_{ij|T|} = 0 \quad \quad \forall i \in \mathcal{I}^{R}, \iota \in \mathcal{N}^i, j \in \mathcal{J}, \label{ConExtended-FreezNoStore}\\
&\sum_{i \in \mathcal{N}} \sum_{\iota \in \mathcal{N}^i} q_i^\iota b_i^\iota (S^{RR\iota}_{ikjt}+S^{RF\iota}_{ikjt}+S^{FR\iota}_{ikjt}+S^{FF\iota}_{ikjt}) \leq C^V_{kj} \quad \quad \forall j,k \in \mathcal{J};j\neq k, t \in \mathcal{T}, \label{ConExtended-ArcCap}\\
&n_j  \leq \sum_{t \in \mathcal{T}}\sum_{\iota \in \mathcal{N}^i}b_i^\iota(x^{R\iota}_{ijt}+x^{F\iota}_{ijt})/a_i \quad \quad \forall i \in \mathcal{I}, j \in \mathcal{J}, \label{ConExtended-FIC}\\
 &Pr \bigg(\sum_{\iota \in \mathcal{N}^i} b_i^\iota(x^{R\iota}_{ijt}+x^{F\iota}_{ijt}) \geq \Delta_{ijt} \bigg)\geq \beta \quad \quad \forall i \in \mathcal{I}, j \in \mathcal{J},t \in \mathcal{T}, \label{ConExtended-DemandLeq-extendedmodel} \\
 & x^{R/F \iota}_{ijt}, n_j, I^{R/F \iota}_{ijt}, S^{RR/RF/FR/FF \iota}_{ikjt}  \geq 0 \quad\quad \forall i \in \mathcal{I}, \iota \in \mathcal{N}^i, j,k \in \mathcal{J};j \not= k, t \in \mathcal{T}.\label{ConExtended_nonnegativity}
\end{align}
\end{subequations}

\end{appendices}
\end{document}